\documentclass{article}
\usepackage{amssymb}
\usepackage{cite,multirow} 
\usepackage[dvips]{graphicx,epsfig}
\usepackage{graphics}
\def\1ad{\mbox{\normalsize $^1$}}
\def\2ad{\mbox{\normalsize $^2$}}
\def\3ad{\mbox{\normalsize $^3$}}
\def\4ad{\mbox{\normalsize $^4$}}
\def\5ad{\mbox{\normalsize $^5$}}
\def\6ad{\mbox{\normalsize $^6$}}
\def\7ad{\mbox{\normalsize $^7$}}
\def\8ad{\mbox{\normalsize $^8$}}

\makeatletter
\@addtoreset{equation}{section}
\makeatother

\newtheorem{meson}{Theorem}[section]

\newcommand{\ignora}[1]{}
\def\beq{\begin{equation}}                     %
\def\eeq{\end{equation}}                       %
\def\bea{\begin{eqnarray}}                     
\def\eea{\end{eqnarray}}                       
\def\0 {\nonumber}

\begin{document}

\setcounter{page}{0}
\begin{titlepage}
\titlepage
\rightline{hep-th/0603253}
\rightline{Bicocca-FT-06-5}
\vskip 3cm
\centerline{{ \bf \Large Deformations of Toric Singularities 
and Fractional Branes}}
\vskip 1cm
\centerline{{\bf Agostino Butti}}
\vskip 1.5truecm
\begin{center}
\em 
Dipartimento di Fisica, Universit\`{a} di Milano-Bicocca \\ 
P.zza della Scienza, 3; I-20126 Milano, Italy\\
\vskip .4cm

\vskip 2.5cm
\end{center}
\begin{abstract}
Fractional branes added to a large stack of D3-branes at the
singularity of a Calabi-Yau cone modify the quiver gauge theory
breaking conformal invariance and leading to different kinds of IR 
behaviors. 
For toric singularities admitting complex deformations we propose a
simple method that allows to compute the anomaly free rank distributions
in the gauge theory corresponding to the fractional deformation
branes.
This algorithm fits Altmann's rule of decomposition of 
the toric diagram into a Minkowski sum of polytopes.
More generally we suggest how different IR behaviors triggered by 
fractional branes can be classified by looking at suitable 
weights associated with the external legs of the (p,q) web.
We check the proposal on many examples and match in some interesting
cases the moduli space of the gauge theory with the deformed geometry.

\vskip1cm

\end{abstract}
\vskip 0.5\baselineskip

\vfill
 \hrule width 5.cm
\vskip 2.mm
{\small
\noindent agostino.butti@mib.infn.it}
\begin{flushleft}
\end{flushleft}
\end{titlepage}
\large
\section{Introduction and overview}

The study of the IR gauge theory on a stack of regular D3 or
fractional branes placed at a Calabi-Yau singularity is an important
issue to test the AdS/CFT correspondence and its extensions to non
conformal cases. 

Many concrete examples has recently been found: the superconformal
gauge theory dual to type IIB string theory on $AdS_5 \times Y^{p,q}$
was built in \cite{benvenuti}; see \cite{kru2,noi,tomorrow} for
$AdS_5 \times L^{p,q,r}$. The Sasaki-Einstein metrics for $Y^{p,q}$
and $L^{p,q,r}$ can be found in \cite{gauntlett} and \cite{CLPP,MSL}
respectively. At the same time many general features of the
correspondence were uncovered, especially for toric Calabi-Yau
singularities:  
the new techniques of dimers, perfect matchings, zig-zag paths 
\cite{dimers,rhombi} allow to represent a complicated
superconformal quiver gauge theory with simple diagrams and to 
compute from them the dual geometry, represented by a toric diagram,
or vice-versa. Therefore it was also possible to perform detailed and
general checks of the correspondence 
\cite{bertolini,hananyX,kru,intriligator,MSY,aZequiv,Tachikawa,Barnes,tri,Lee}.
Alternative techniques for the study of Calabi-Yau singularities are
based on exceptional collections \cite{Herzog,exceptional}.

A well known method to break conformal invariance is to add
fractional branes, that can be seen as higher dimensional branes
wrapping collapsed cycles at the singularity. On the gauge theory side
the fractional branes modify the number of colors of different gauge
groups consistently with cancellation of anomalies for gauge
symmetries.
In many known examples \cite{KS,cascateypq,multiflux} 
fractional branes lead to cascades of Seiberg
dualities that reduce the number $N$ of regular branes, so that the IR
dynamics is dominated by fractional branes.

A classification of fractional branes into three different classes
according to the IR behavior they produce in the gauge theory was
proposed in \cite{susyb1}. We may have i) fractional deformation
branes, that describe a complex deformation of the dual geometry and
produce a supersymmetric (typically confining) vacuum in field theory;
ii) $\mathcal N=2$ fractional branes, leading to $\mathcal N=2$
dynamics in some regions of the moduli space of the gauge theory and
 iii) supersymmetry breaking (SB)
fractional branes, that seem to be the most common kind of fractional
branes: a supersymmetric vacuum is no more present and typically one
finds a runaway behavior \cite{susyb1,susyb2,susyb3,runaway,force}. 
   
In general there is a great number of fractional branes that can be
consistently added to a quiver gauge theory: in the toric case there
are $d-3$ fractional branes, where $d$ is the perimeter of the toric
diagram of the dual geometry. Therefore one would need a simple method
to compute the anomaly free rank distributions corresponding to the three 
classes of fractional branes. In this paper we propose an
algorithm to do that in the general toric case. We will use the
language of dimers and zig-zag paths.

First of all we use the known correspondence between fractional branes,
that is anomaly free rank distributions in the gauge theory, and the $d-3$ 
baryonic symmetries of the original superconformal theory (without
fractional branes) \cite{phases}. Then the main idea of this paper is to
parametrize the global symmetries, and among them the baryonic
symmetries, using weights $b_i$ assigned to the external legs $v_i$ of the 
(p,q) web, or equivalently to the zig-zag paths in the dimer
configuration. The global charge of any link in the dimer is computed by the
difference of the two weights of the zig-zag paths to which the link
belongs. 

It is then easy to understand to which class of fractional branes 
a rank distribution in the gauge theory belongs by looking at
the weights of the corresponding baryonic symmetry. We will treat 
in great detail the case of deformation branes: even though the 
deformation of a toric Calabi-Yau cone is no more a toric manifold,
having only $U(1)^2$ isometries, there is a simple rule based only
on toric data to understand whether a toric cone admits a complex 
structure deformation. In fact deformations of isolated Gorenstein
singularities are in correspondence with Minkowski decompositions
of the toric diagram \cite{altmann} or equivalently with splittings
of the (p,q) web into sub-webs in equilibrium. Our proposal is that
fractional branes corresponding to such deformed geometries have 
constant weights $b_i$ on the different sub-webs.

$\mathcal N=2$ fractional branes instead are possible when there is
a not isolated singularity, that is when in the (p,q) web there are 
parallel vectors perpendicular to the same edge of the toric diagram.
In our proposal the baryonic symmetries associated with $\mathcal N
=2$ fractional branes have non-zero weights only on these parallel
vectors. 

We also suggest that different assignments of weights $b_i$ correspond
to SB fractional branes.     

We check these proposals on concrete examples. In particular for theories
admitting complex deformations, when a single deformation parameter 
is turned on, we show that gauge groups have the only possible ranks:
$SU(N)$, $SU(N+M)$, $SU(N-M)$ (previously in the literature only cases
with $SU(N)$ and $SU(N+M)$ gauge groups were known); moreover in these 
cases our proposal for deformation branes leads to configurations where
no gauge group can develop an ADS superpotential term, and therefore
the existence of a supersymmetric vacuum is expected. In this analysis
we also use the splitting into sub-webs at the level of zig-zag paths in
the dimer, that has already been observed in a recent paper \cite{uranga}.

If we add fractional deformation branes we should not only check that our
proposal leads to a supersymmetric vacuum, but also that the quantum
modified moduli space of the gauge theory, when probed by a regular
brane, is equal to the complex deformation 
of the toric singularity. For some examples of such computations in
the literature see 
\cite{KS,multiflux}; interesting are the techniques used in \cite{Pinansky}, 
since they should work for all toric cases admitting deformations. 
One has to write the moduli space of vacua through F-term relations in
the chiral ring of mesonic operators, that are typically modified at 
quantum level by ADS terms; on the geometric side the linear relations in $C^*$
(the dual of the toric fan $C$) expressing the toric manifold as a 
(non complete) intersection in a complex space can be modified using 
Altmann's results \cite{altmann}. 

We study the example of the $PdP_4$ theory, admitting
two complex deformation parameters, in order to verify that our proposal
for computing the rank distribution for fractional deformation branes 
reproduces correctly also the deformed geometry. 
In performing these computations we make use of the $\Psi$-map,
recently introduced in \cite{exceptional}, since it
allows to find the precise mapping between mesons in the chiral ring and
integer points in $C^*$, as already noted in the same paper. 

Therefore we translate the $\Psi$-map theory in \cite{exceptional}
in the language of charges and zig-zag paths. We also note that the
idea of giving weights to zig-zag paths allows to prove explicitly that
the $\Psi$-map of a closed loop is an affine function and that the
flavor charges of mesons are proportional to the homotopy numbers of the
corresponding loops in the torus, as observed for the first time in
\cite{kru}.   
 
This paper is organized as follows. Section \ref{gen} contains the
definitions of useful tools like dimers, zig-zag paths and the
algorithms for distributing charges in the dimer \cite{aZequiv,proc}. 
In Section
\ref{classes} we review the classification of fractional branes
according to the IR behavior
\cite{susyb1} and the Minkowski decomposition of toric
diagrams. In Section \ref{fractional} we explain in great detail the
correspondence between anomaly free rank distributions in the dimer
and baryonic symmetries of the superconformal theory \cite{phases}; we
also prove that the correspondence is one to one. In Section
\ref{matching} we introduce the parametrization of  global charges through
weights for zig-zag paths and we characterize the
three classes of fractional branes through the weights of the associated
baryonic symmetries. We check this proposal for computing rank
distributions in many examples in Section \ref{check}, where we also
treat the general case of theories with a single deformation parameter
turned on.
Section \ref{psimap} contains useful comments to the $\Psi$-map theory
\cite{exceptional}. In Section
\ref{example} we provide the explicit computation for $PdP_4$ of the
moduli space of the gauge theory with fractional branes that matches
the deformed geometry.   

\section{Generalities about the gauge theory}
\label{gen}

In this Section we briefly review some results about the AdS/CFT
correspondence in the superconformal case that have been recently
obtained for toric geometries. To be concrete we will explain the
ideas on a specific example well known in the literature: the
Suspended Pinch Point (SPP) that we will use also in the following
Sections. 

We consider $N$ D3-branes living at the tip of a CY cone.
The base of the cone, or horizon, is a five-dimensional
compact Sasaki-Einstein manifold $H$ \cite{kw,horizon}.
The IR limit of the gauge theory living on the branes is 
${\cal N}=1$ superconformal and dual in the AdS/CFT correspondence 
to the type IIB background $AdS_5\times H$, which is 
the near horizon geometry.

The problem of finding the low energy gauge theory dual to a generic
Calabi-Yau singularity is difficult and still unsolved, but recently 
the AdS/CFT correspondence has been built for a wide class of CY
singularities: the toric CY cones (roughly speaking a six dimensional
manifold is toric if it has at least $U(1)^3$ isometries). 

\begin{figure}
\begin{center}
\includegraphics[scale=0.6]{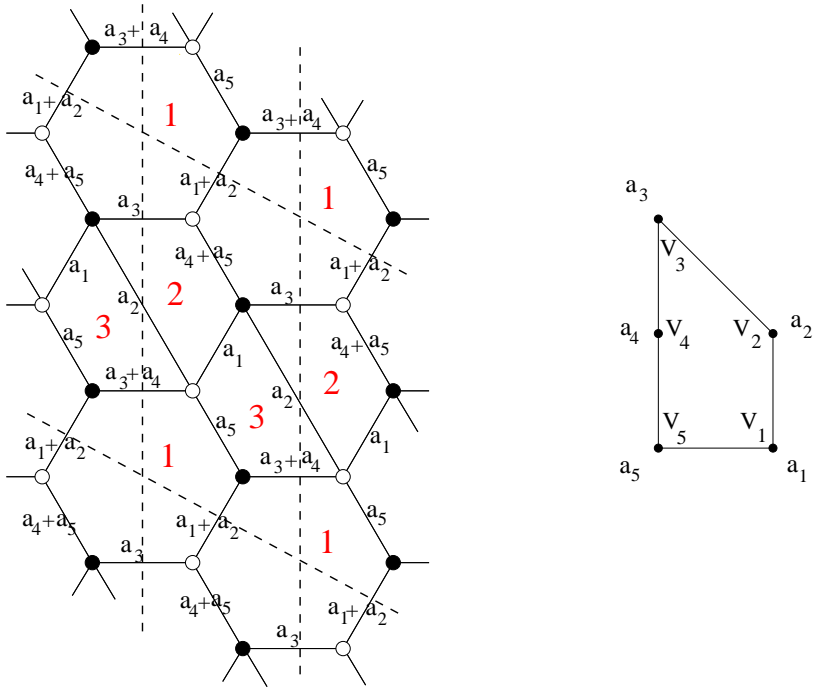} 
\caption{Dimer configuration and toric diagram for the Suspended Pinch
Point.}
\label{spp1}
\end{center}
\end{figure}

\begin{figure}
\begin{center}
\includegraphics[scale=0.6]{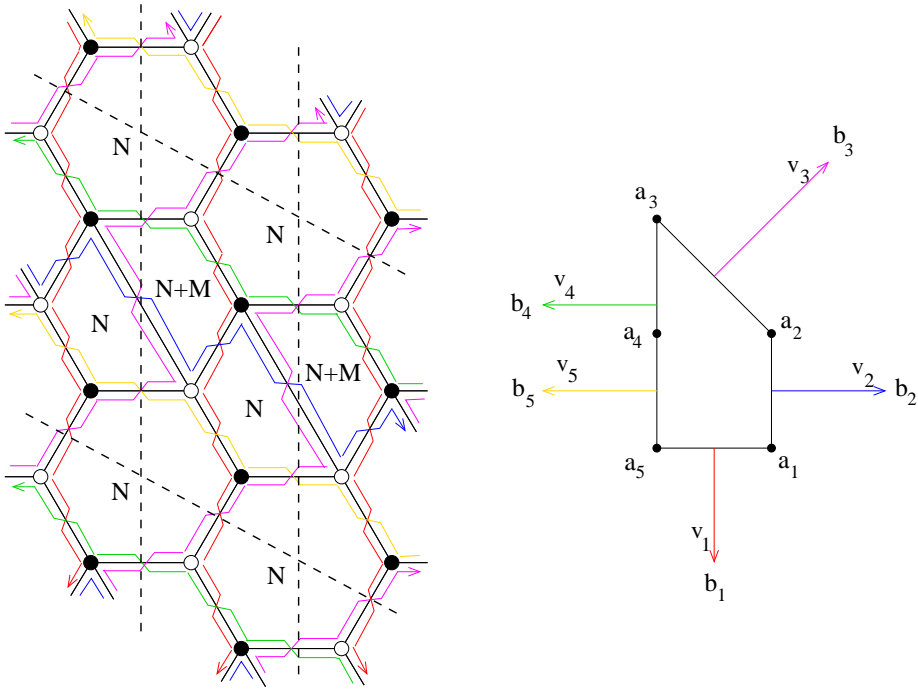} 
\caption{Zig-zag paths for the Suspended Pinch Point and their
  correspondence with external legs of the (p,q) web.}
\label{spp2}
\end{center}
\end{figure}

Many geometrical informations about toric CY cones are encoded in
the \emph{toric diagram}, a convex polygon in the plane with integer
vertices. For the SPP example the fan $C$ is generated by the integer
vectors $V_i$:\footnote{Because of the CY
condition it is always possible to choose the third coordinate of
the vectors $V_{i}$ equal to $z=1$. The toric diagram is the
intersection of the fan with the plane $z=1$. For an introduction to
toric geometry see \cite{fulton} and the review part of \cite{MS}.} 
\begin{equation}
(0,0,1) \quad (1,0,1) \quad (1,1,1) \quad (0,2,1)
\end{equation}
and the corresponding toric diagram is drawn in Figure \ref{spp1}.
The \emph{(p,q) web} is the set of vectors perpendicular to the edges of the 
toric diagram and with the same length as the corresponding edges (see
Figure \ref{spp2}).

In the toric case the gauge theory is completely identified by the
\emph{periodic quiver}, a diagram drawn on $T^2$ (it is the ``lift'' 
of the usual
quiver to the torus): nodes represent $SU(N)$ gauge groups, oriented
links represent chiral bifundamental multiplets and faces represent
the superpotential: the trace of the
product of chiral fields of a face gives a superpotential 
term (with a sign + or - if the arrows of the face in the
periodic quiver are oriented clockwise or anticlockwise respectively). 

Equivalently the gauge theory is described by the
\emph{dimer configuration}, or \emph{brane tiling},
the dual graph of the periodic quiver, drawn also on a torus $T^2$.
In the dimer the role of faces and vertices is exchanged: 
faces are gauge groups and vertices are superpotential terms.
The dimer is a bipartite graph: it has an equal number of white and 
black vertices (superpotential terms with sign + or - respectively)  
and links connect only vertices of different colors.

The dimer for SPP is drawn in Figure \ref{spp1}: it has three faces
$F=3$, seven edges $E=7$, and four vertices $V=4$. 
The three gauge groups are labelled by
the red numbers in Figure \ref{spp1}: faces with the same number are
identified. The fundamental cell of the torus $T^2$ where the dimer
lies is (any of) the parallelogram formed by the dashed lines. Since
the dimer is on a torus we have: $V-E+F=0$.
   
By applying Seiberg dualities to a quiver gauge theory we can obtain
different quivers that flow in the IR to the same CFT: to a toric
diagram we can associate different quivers/dimers describing the same
physics. It turns out that one can always find phases where all the 
gauge groups have the same number of colors; these are called 
\emph{toric phases}. Seiberg dualities keep constant the number of
gauge groups $F$, but may change the number of fields $E$, and
therefore the number of superpotential terms $V=E-F$. We will call 
\emph{minimal toric phases} those having the minimal number of fields
$E$. 

If the dimer is known the toric diagram can be reconstructed using
\emph{perfect matchings}. A perfect matching is a subset of links in the
dimer such that every white and black vertex is taken exactly
once. Perfect matchings can be mapped to integer points of the toric
diagram through the Kasteleyn matrix which counts their (oriented)
intersections with two loops generating the fundamental group of the
torus \cite{dimers}.

The inverse problem of reconstructing the dimer from the toric diagram
can be solved using \emph{zig-zag paths} \cite{rhombi} (see also
\cite{mirror}). 
A zig-zag path in the dimer is a path of links that turn maximally 
left at a node, maximally right at the next node, then again 
maximally left and so on \cite{rhombi}. 
We draw them in the specific case of SPP theory in Figure \ref{spp2}: 
they are the five loops in red, blue, magenta, green and yellow and 
they are drawn so that they intersect in the middle of a link as in 
\cite{mirror}. Note that every link of the dimer belongs to exactly 
two different zig-zag paths, oriented in opposite directions. 
Moreover for dimers representing consistent theories the zig-zag paths 
are closed non-intersecting loops. There is a \emph{one to one
correspondence between zig-zag paths and legs of the $(p,q)$ web}: 
the homotopy class in the fundamental group of the torus of every 
zig-zag path is given by the integer numbers $(p,q)$ of the 
corresponding leg in the $(p,q)$ web \cite{rhombi}. The reader can
check this in the example of Figure \ref{spp2}. Note that
there are two distinct zig-zag paths with homotopy numbers (-1,0) and
not a unique path with homotopy (-2,0) that would intersect
itself. This is a general feature of theories with a toric diagram
having integer points on its edges.  

The \emph{Fast Inverse Algorithm} of \cite{rhombi} consists just in drawing 
the zig-zag paths on a fundamental cell with the appropriate 
homotopy numbers and satisfying suitable consistency conditions.

\subsection{Distribution of charges in the dimer}
\label{distribution}

Non anomalous $U(1)$ symmetries play a very important role in the
gauge theory. Here we review how to count and parametrize them and
how to compute the charge of a certain link in the dimer.

For smooth horizons $H$ we expect $d-1$ global non anomalous symmetries, 
where $d$ is the number of sides of the toric diagram in the dual theory.
We can count these symmetries from the 
number of massless vectors in the $AdS$ dual. Since the manifold is toric, 
the metric has three $U(1)$ isometries.
One of these (generated by the Reeb vector) corresponds to the
R-symmetry while the other two give two global flavor symmetries 
in the gauge theory. Other gauge fields in $AdS$ come 
from the reduction of the RR four form on the non-trivial three-cycles
in the horizon manifold $H$, and there are $d-3$ three-cycles in
homology \cite{tomorrow}  when $H$ is smooth.
On the field theory side, these gauge fields  correspond to baryonic 
symmetries. Summarizing, the global non anomalous symmetries are:
\begin{equation}
U(1)^{d-1}=U(1)^2_F \times U(1)^{d-3}_B
\label{count}
\end{equation}
If the horizon H is not smooth (that is the toric diagram has integer
points lying on the edges), equation (\ref{count}) is still true with $d$
equal to the perimeter of the toric diagram in the sense of toric
geometry (d = number of vertices of toric diagram + number of integer
points along edges). For instance in the SPP theory $d=5$, so that
there are 2 baryonic symmetries.

These $d-1$ global non anomalous charges can be parametrized by $d$
parameters $a_1, a_2, \ldots ,a_d$ \cite{aZequiv}\footnote{The
algorithm proposed in \cite{aZequiv} to extract the field theory
content from the toric diagram is a generalization of previously known
results, see for instance \cite{benvenuti,tomorrow,hananymirror}, 
and in particular of the folded quiver in \cite{kru2}.}, each associated 
with a vertex of the toric diagram or a point along an edge 
(see Figure \ref{spp2} for SPP), satisfying the constraint:
\begin{equation}
\sum_{i=1}^d a_i = 0
\label{sum}
\end{equation}
The $d-3$ baryonic charges are those satisfying the further
constraints \cite{tomorrow}:
\begin{equation}
\sum_{i=1}^d a_i V_i=0
\label{bar}
\end{equation}
where $V_i$ are the vectors of the fan: $V_i=(x_i,y_i,1)$ with
$(x_i,y_i)$ the coordinates of integer points along the perimeter of
the toric diagram. 

As an aside recall that R-symmetries are parametrized with the $a_i$
having total sum 2 instead of zero.

There are two simple equivalent algorithms to compute the charge of a
generic link in the dimer in function of the parameters $a_i$
(the equivalence of the two algorithms was shown in \cite{proc}, 
assuming a conjecture in \cite{rhombi}).

The first efficient way to find the distribution of charges \cite{aZequiv}, 
valid for all toric phases, is based on perfect matchings: 
the parameters $a_i$ are associated with vertices of the toric
diagram, and to every 
vertex $V_i$ there corresponds a single perfect matching in the dimer, 
at least for physical theories \cite{aZequiv,rhombi}. Therefore 
the charge of a link in the dimer can be computed as the sum of 
the parameters $a_i$ of all the external perfect matchings 
(corresponding to vertices) to which the link belongs. For examples
of how to use this prescription using the Kasteleyn matrix, 
see \cite{aZequiv}.
 
The second algorithm is based on zig-zag paths \cite{proc}. 
Consider the two zig-zag paths to which a link in the dimer belongs. 
They correspond to two vectors $v_i=(p_i, q_i)$ and $v_j=(p_j, q_j)$ 
in the $(p,q)$ web. Then the charge of the link is given by the sum 
of the parameters $a_{i+1}+a_{i+2} \ldots +a_{j}$ between the vectors
$v_i$ and $v_j$ \footnote{For minimal toric phases it is always
possible to choose the sum of the parameters $a_i$ in the angle less than
$180^o$ formed by $v_i$ and $v_j$. We will generalize this to all
toric phases in Section \ref{matching}.}.  
So for instance in Figure \ref{spp2} the links corresponding 
to the intersection of the red and the magenta zig-zag paths 
(vectors $v_1$ and $v_3$ in the (p,q) web) have charge equal to $a_1+a_2$. 

This rule explains the formula for the multiplicities of fields 
with a given charge \cite{aZequiv}: 
since every link in the dimer corresponds to the intersection of two
zig-zag paths, the number of fields with charge $a_{i+1}+a_{i+2} \ldots +a_{j}$
is equal\footnote{This is true in minimal toric phases, where the
  number of real intersections between two zig-zag paths is equal to
  the topological number of intersections.} to the number of 
intersections between the zig zag paths 
corresponding to $v_i$ and $v_j$, which is just $\mathrm{det}(v_i,v_j)$.

\section{Classes of fractional branes and deformations \\ in toric geometry}
\label{classes}

In this Section we review the classification \cite{susyb1} of the 
different types
of IR behaviors that fractional branes can induce in the gauge theory.
We also explain
Altmann's rule for understanding deformations of toric singularities
\cite{altmann} . 

Let us start with a large number $N$ of regular $D3$-branes at the
singularity of a toric CY cone. The IR limit of the gauge theory on
the branes is superconformal and the dual geometry is $AdS_5 \times
H$. A well known method to break the conformal symmetry is to add
fractional branes.
Fractional branes may be
thought as higher dimensional branes wrapping collapsed cycles at the
singularity. From the dual string theory point of view they add new
fluxes and change the $AdS$ geometry \cite{KS,cascateypq}: 
the only known example of
smooth metric describing the near horizon geometry produced by
fractional branes is the Klebanov-Strassler solution \cite{KS}, relative
to fractional branes at the tip of the conifold ($H=T^{1,1}$); the
internal metric is the CY metric over the deformed conifold up to a
warp factor.

It is easier to study fractional branes from the gauge theory
point of view: in this case they are described by a
modification of the number of colors of the different gauge groups in 
the quiver gauge theory (with the only requirement that gauge
symmetries are still anomaly free). Maybe it is simpler to see this on
the mirror description \cite{mirror}: the mirror of the apex of the cone
(where the $T^3$ fibration of the toric manifold is completely
degenerate) is a ``pinched'' $T^3$ made up of a collections of $F$
intersecting $S^3$, where $F$ is the number of gauge groups. 
D3 regular, and D5, D7 fractional branes at the
singularity are mapped in the mirror to D6-branes wrapping the $S^3$'s: in
particular the $N$ D3 branes are mapped to D6-branes wrapping all the 
$S^3$'s, contributing to the same factor of $N$ to the number of
colors of gauge groups, whereas fractional branes 
wrap only some of the $S^3$, modifying the rank distribution.

In this paper when speaking of fractional branes we will always refer
to this changing of rank distribution in the quiver gauge theory.

The IR behavior of quiver gauge theories with fractional branes have
been studied in many examples, for recent works see
\cite{multiflux,susyb1,susyb2,susyb3,uranga}. In \cite{susyb1} a general
classification of fractional branes was accordingly proposed:
\begin{itemize}
\item{\bf Deformation fractional branes:} They are present when the
  dual toric geometry admits a complex structure deformation according
  to Altmann's rule and they describe the gauge theory dual to the
  geometry of the deformed cone. 
  In fact when there is no obstruction to a deformation,
  we may expect the existence of a CY
  metric over the deformed cone and a supergravity solution similar to
  the Klebanov-Strassler solution. 
  The gauge theory has therefore a supersymmetric vacuum ($\mathcal
  N=1$). In concrete
  examples it turns out that these fractional branes lead to a
  cascading behavior: after a certain number of Seiberg dualities the
  gauge theory comes back to itself with the number of regular branes
  decreased by some amount of fractional branes $N \rightarrow N-M$. 
  Typically in the IR one finds a certain number
  of isolated confining gauge groups with no more bifundamental matter.    
\item{\bf $\mathcal N=2$ fractional branes:} They are present when
  there are integer points along the sides of a toric diagram. In this
  case the horizon $H$ is not smooth and the singularity at the tip of
  the cone is not isolated. A side of the toric diagram with $k-1$
  internal points gives rise to a complex line of $\mathbb
  C^2/\mathbb Z_k$ singularities in the toric cone passing through the
  origin, along which the fractional branes can move. The dual gauge
  theory has flat directions where the dynamics has an
  accidental $\mathcal N=2$ supersymmetry. The behavior is really
  different from the case of deformation branes, for example the SPP
  theory with an $\mathcal N=2$ fractional brane, corresponding to
  Figure \ref{sppcharges} c), does not seem to have
  a cascade. 
\item{\bf Supersymmetry breaking (SB) fractional branes:} 
  They seem to be the most generic
  anomaly free distributions of ranks for gauge groups; they are present
  also when the toric cone admits a deformation, in fact the number of
  complex structure deformations is less (or at most equal, see the KS
  theory) than the number of possible
  fractional branes (= number of baryonic symmetries in the original
  superconformal theory = $d-3$).
  Their main feature is that the gauge theory does not have a 
  supersymmetric vacuum\cite{susyb1,susyb2,susyb3,runaway,force}: 
  typically in the IR, when the number of colors
  has decreased after the cascade, some gauge group have
  $N_f<N_c$ and develops ADS superpotential terms leading to a runaway
  behavior.
\end{itemize} 

\begin{figure}
\begin{center}
\includegraphics[scale=0.50]{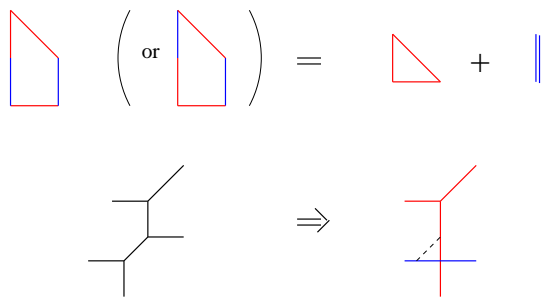}
\caption{Deformation of the cone over SPP: decomposition of the toric
diagram in Minkowski summands and splitting of the web into subwebs
in equilibrium.}
\label{defspp}
\end{center}
\end{figure}

In Section \ref{matching} we will explain how to find the rank
distributions corresponding to these kinds of fractional branes.

Let us now briefly explain Altmann's rule for deformations of toric
singularities. In \cite{altmann} it is shown that the complex
deformations of isolated Gorenstein (i.e. CY) toric singularities are
completely characterized by the possible decompositions of the toric
diagram into a Minkowski sum of polytopes.
We will deal the case of 6d toric cones, described by toric diagrams
on a 2-plane. 
Given two 2d convex polygons $P_1$, $P_2$, one can define their Minkowski sum
$P_1+P_2$ as the convex hull of the set $\{p=p_1+p_2 | \, p_1 \in P_1, p_2
  \in P_2\}$, that is the set of points obtained by summing the points
of the two polygons. One can realize that the edges of $P$ are the
union of the sets of edges of the polygons $P_1$ and $P_2$.

We give an example in Figure \ref{defspp}, where we show the
decomposition of the toric diagram of SPP into two Minkowski summands
(note however that this singularity is not isolated, and hence one can
identify the sides of the triangle in red in two different ways into
the original SPP toric diagram, compare with Figures \ref{sppcharges}
a) and b)). 

It is possible to read the same decomposition in terms of
subdivision of the (p,q) web into two (or more) sub-webs at
equilibrium, that is the perpendiculars to the sides of the toric
diagram are divided into subsets where the sum of vectors is still
zero. We show this in the same Figure \ref{defspp}: the two legs in
blue are lifted from the plane of the other vectors and the link
between the two subwebs represents a three-cycle (one deformation
parameter). Therefore the cone over SPP has $d-3=2$ fractional branes
and a branch of complex deformations with one parameter.

\begin{figure}
\begin{center}
\includegraphics[scale=0.50]{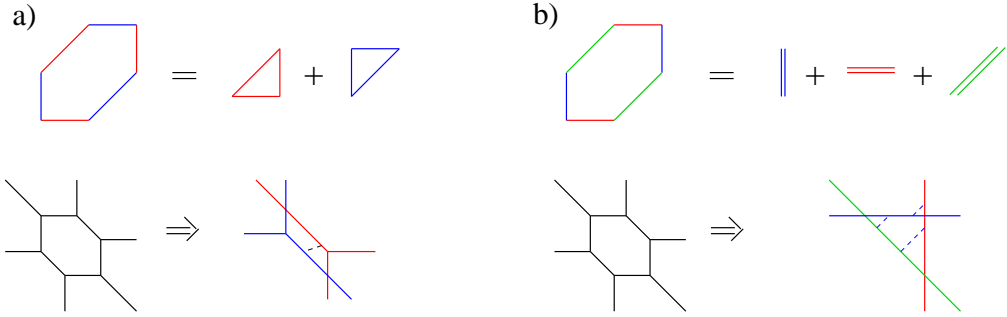}
\caption{Deformation of the cone over $dP_3$. a) One parameter
  branch. b) Two parameters branch.}
\label{defdp3}
\end{center}
\end{figure}

In Figure \ref{defdp3} we report the example of the cone over $dP_3$.
We see that there are two possible decompositions into Minkowski
summands, that is two branches of complex structure deformations: the
first one, Figure \ref{defdp3} a), has one parameter (separation into 
two sub-webs). The second branch, Figure \ref{defdp3} b) has
two parameters (separation into three subwebs). The number of
fractional branes for this theory is $d-3=3$.

Toric cones whose toric diagram has no Minkowski decompositions do not
admit complex structure deformations.
  
\section{Fractional branes and baryonic symmetries}
\label{fractional}

In this Section we explain in detail the known correspondence
between fractional branes and baryonic symmetries in the gauge theory
\cite{phases}. We also prove that the correspondence is one to one.

As explained in the previous Section, fractional branes in the 
quiver gauge theory
modify the number of colors of the gauge groups in such a way 
that the gauge symmetries are still anomaly free; 
in fact the number of flavors for every gauge group
is equal to the number of anti-flavors. 
These anomaly free configurations can be computed through
the (integer) kernel of the antisymmetric intersection matrix
$S_{ij}$ defining the quiver:
\begin{equation}
\sum_j S_{ij} \, n_j =0
\label{kernel}
\end{equation}
where in this Section the indexes $i,j$ label the gauge groups, that is the
nodes of the periodic quiver. $n_j$ is the number of colors of
j-th gauge group, and the entry $(i,j)$ of the intersection
matrix $S_{ij}$ is the number of arrows going from node $i$ 
to $j$ minus the number of arrows going from $j$ to $i$.
In the following we will assume to start from a toric phase of
the original superconformal gauge theory (before introducing 
fractional branes), that is a phase where all the gauge 
groups have the same number of colors $SU(N)$. 
Therefore the constant vector $n_j=N$ is always in the 
kernel of $S$ and it describes regular D3-branes. 

There is an equivalent way to compute the allowed distributions 
of colors $n_i$ based on baryonic symmetries. 
Consider a rank assignment $n_i$ as in (\ref{kernel}). 
Define for every oriented link $X$ of the periodic quiver a charge $c(X)$:
\begin{equation}
c(X_{i \rightarrow j})=n_j-n_i
\label{char}
\end{equation}
if the field $X$ goes from node $i$ to node $j$. Note that 
adding regular branes does not change the charges of chiral fields.

The charges in (\ref{char}) can be seen as 
linear combinations of the $U(1)$ parts of the original 
$U(N)$ gauge groups: the charge associated with the i-th 
$U(1)$ gauge group is $+1$ ($-1$) for links entering (exiting) 
in the i-th node and zero for other fields. 
The sum of these charges with weight $n_i$ for each node $i$
gives the distribution in (\ref{char}). 

It is easy to see that (\ref{char}) defines a global non 
anomalous (baryonic) $U(1)$ charge of the original superconformal 
gauge theory, that is the theory with all groups equal to $SU(N)$. 
First of all note that from (\ref{char}) it follows that the total 
charge of every closed loop of links in the periodic quiver 
is zero (this is true also if the arrows are not all oriented 
in the same direction: we simply define the charge of a loop 
by subtracting the charges of links oriented in the opposite
direction). 
In particular faces of the periodic quiver are closed loops 
and represent superpotential terms, and hence the superpotential 
is conserved.

To check that the symmetry is non anomalous under every 
gauge transformation we have to compute for every gauge 
group $i$ the sum of the charges of all links attached to node $i$; 
this is given by:
\begin{equation}
\sum_{j} S_{ij}\, c(X_{i \rightarrow j})= 
\sum_{j} S_{ij}\,n_j -\sum_{j} S_{ij}\,n_i=0
\label{equa}
\end{equation}
which vanishes because of (\ref{kernel}) and 
because the original phase is toric.

Vice versa every global non anomalous $U(1)$ symmetry 
(with integer coefficients) such that every closed loop 
(oriented or not) has charge zero defines a rank 
assignment satisfying (\ref{kernel}): start from 
a generic gauge group $i$ and fix its rank to 
an arbitrary integer $n_i$. Then the rank of a 
node $j$ connected to $i$ by a path $L$ is obtained as:
\begin{equation}
n_j=n_i+c(L_{i \rightarrow j})
\label{build}
\end{equation}
Since the charge of closed loops is zero, this rank 
assignment is unambiguous. Moreover from the fact that 
the $U(1)$ symmetry is non anomalous in the original 
superconformal theory, that is 
$\sum_{j} S_{ij}\, c(X_{i \rightarrow j})=0$, 
we find that equation (\ref{kernel}) is satisfied 
(look at the first equality in (\ref{equa})). Note that 
all ranks are defined up to a common constant, 
that can be varied by adding regular branes. Equation (\ref{build}) or
(\ref{char}) also shows that global $U(1)$ symmetries that assign zero
charge to all closed (oriented or not) loops are automatically linear
combinations of the $U(1)$ parts of the original $U(N)$ gauge groups.

Therefore we have a one to one correspondence 
between fractional branes (\ref{kernel}) and non anomalous global $U(1)$ 
symmetries that assign zero charge to all closed loops. It is 
known in the literature that such symmetries are 
the baryonic symmetries (of the theory with all 
$SU(N)$ gauge groups). As an evidence for this recall 
that mesonic operators in the superconformal field 
theory are dual to supergravity states in string theory, 
whereas baryonic operators, having a conformal dimension 
proportional to $N$, correspond to states of a D3-brane 
wrapped over opportune three cycles of the horizon manifold 
$H$. Therefore only baryons can be charged under a 
baryonic symmetry, that in the string theory dual comes from 
the reduction of RR four form along three cycles in $H$. 
Instead mesonic operators, that are closed oriented loops, have 
zero charge under baryonic symmetries.

In Section \ref{psimap}, we will give a direct proof in the gauge 
theory for the toric case that 
the $d-3$ baryonic symmetries are exactly the symmetries 
under which all loops (also non  oriented) have zero charge. Instead
the charges of loops under the two flavor symmetries are proportional
to the homotopy numbers of the loops in the torus $T^2$ where the
periodic quiver is drawn.

\section{Matching deformations with fractional branes}
\label{matching}

In this Section we propose a simple method to find the rank distribution of
gauge groups in the gauge theory dual to the geometry produced by
fractional deformation branes, when the toric
singularity admits a complex-structure deformation according to
Altmann's rule.  We will also extend the proposal
to $\mathcal N=2$ branes.

First of all we have to find the baryonic symmetry associated with 
the fractional
brane and then reconstruct the ranks of the gauge groups
as explained in the previous Section, equation (\ref{build}).

\begin{figure}
\begin{center}
\includegraphics[scale=0.67]{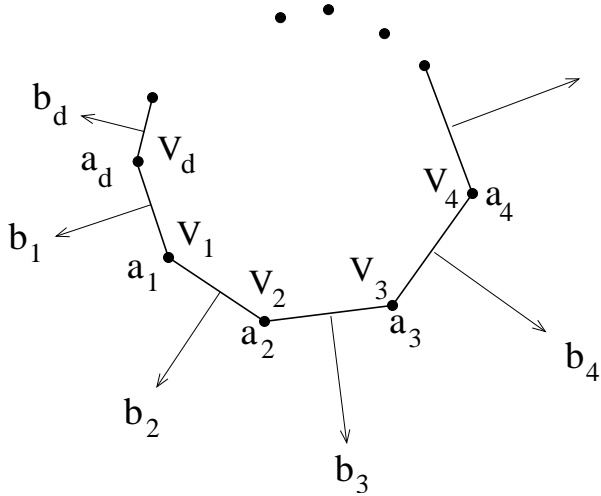}
\caption{The parameters $b_i$ for global charges.}
\label{weights}
\end{center}
\end{figure}

The main idea is to change the parametrization of global charges: 
instead of using the parameters $a_i$ associated with integer points 
on the boundary of the toric diagram and satisfying equation 
(\ref{sum}), we introduce new parameters $b_i$ associated 
with the vectors $v_i$ of the (p,q) web, that are the 
perpendiculars to the edges of the toric diagram, 
see Figure \ref{weights}. In this Section $i=1,\ldots d$ labels legs
of the (p,q) web or integer points along the boundary of the toric
diagram. $d$ is the perimeter of the toric diagram. 

For a global charge, the new parameters $b_i$ are determined 
so that they satisfy the relations:
\begin{equation}
a_i=b_{i+1}-b_{i} \qquad \quad \forall i=1, \ldots d
\label{relation}
\end{equation}
and this is possible because of equation (\ref{sum}). Moreover 
the $b_i$ are defined up to an additive common constant, 
that leaves unchanged the $a_i$, so that we get a parametrization 
of the $d-1$ global charges. Note that in our conventions the 
$a_i$ and $b_i$ are distributed anticlockwise along the toric 
diagram or (p,q) web and $a_i$ is placed between the legs with 
parameters $b_i$ and $b_{i+1}$ as in Figure \ref{weights}. 
The indexes $i$ are understood to be periodic with period $d$.

Equation (\ref{relation}) implies analogous relations for the 
charges of ``composite'' fields: for example the field with 
charge $a_1+a_2$ in the SPP example can be reparametrized as 
$b_3-b_1=(b_3-b_2)+(b_2-b_1)$. And note in Figure \ref{spp2} 
that this field is just the intersection of the zig-zag paths 
corresponding to the vectors $v_1$ and $v_3$ in the (p,q) web, 
in agreement with the algorithm for distributing charges 
proposed in \cite{proc}.
In fact because of the correspondence between vectors of the 
(p,q) web and zig-zag paths, we can think that the weights 
$b_i$ are assigned to zig-zag paths in the dimer.

\begin{figure}
\begin{center}
\includegraphics[scale=0.67]{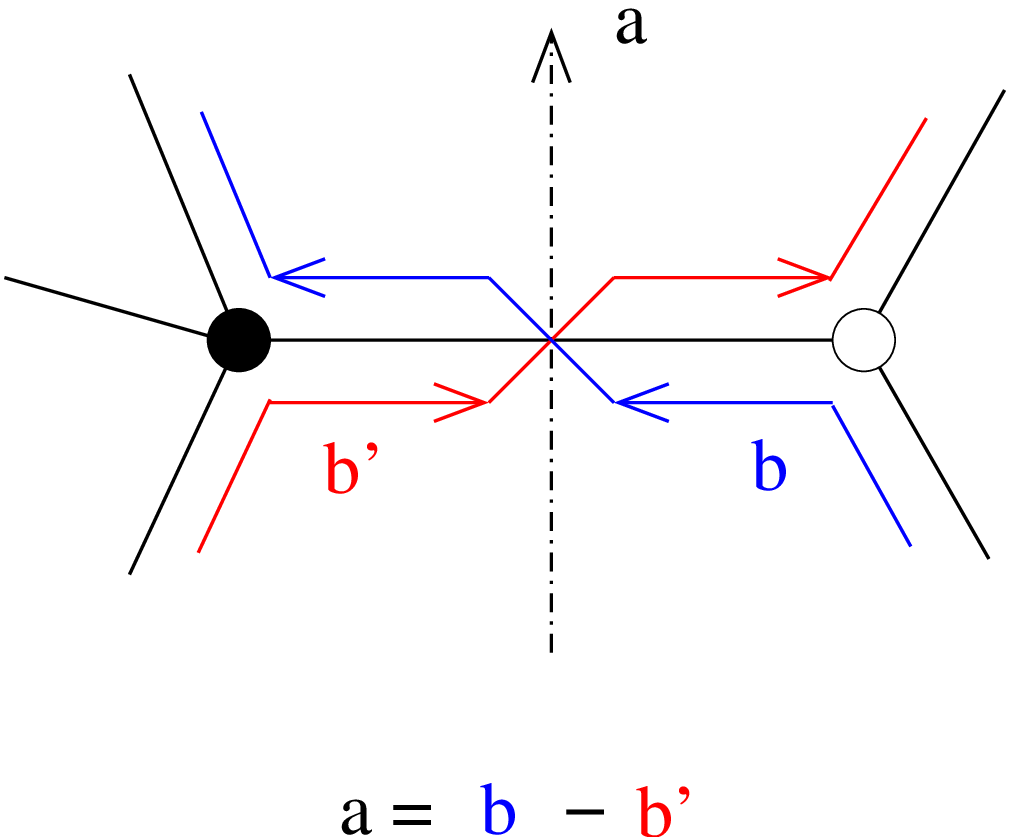}
\caption{The charge of a link in the dimer in function of the 
weights $b$ of the zig-zag paths.}
\label{charge}
\end{center}
\end{figure}

Let us restate more precisely the method to find the global 
charge of a link in the dimer in functions of the parameters $b_i$. 
Look at Figure \ref{charge}: we orient the chiral field in the 
periodic quiver so that the white vertex of the dimer is on the right. 
With respect to this orientation of the chiral field the two 
zig-zag paths defining the link always arrive from the bottom 
and go out from the top of the link in the dimer. This is 
because the zig-zag paths always turn clockwise around white 
nodes and anticlockwise around black nodes (this is a consistency rule 
for the Fast Inverse Algorithm \cite{rhombi}). If $b$ is the weight 
of the zig-zag path entering at bottom right and going out at 
top left, and $b'$ is the weight of the other zig-zag path 
entering at bottom left and going out at top right, then the 
global charge of the corresponding chiral field is always:
\begin{equation}
a=b-b'
\label{carica}
\end{equation}
This is a precise reformulation of the algorithm in \cite{proc} 
that can be extended without ambiguities to all toric phases.

Note also that it is immediate to prove that rule (\ref{carica}) 
gives global non anomalous charges. The sum of the charges of 
the chiral fields connected to a node in the dimer is zero 
(invariance of the superpotential) since every zig-zag path appears 
twice in consecutive links, but its weight $b$ is once added and 
once subtracted. For the same reason the sum of global charges 
of links for every face of the dimer is zero (anomaly cancellation).  

To find the baryonic charges we have to impose the constraint
(\ref{bar}):
\begin{equation}
0=\sum_i a_i V_i= \sum_i b_{i+1} V_i -b_i V_i = 
\sum_i b_i(V_{i-1}-V_{i})
\end{equation}
and since the difference $(V_{i-1}-V_{i})$ of consecutive vectors
in the fan is proportional up to a rotation of $90^o$ to the vector
$v_i$ of the (p,q) web, we find that the $d-3$ baryonic charges are
those satisfying the constraints:
\begin{equation}
\sum_{i=1}^d b_i v_i =0
\label{bar2}
\end{equation} 
Note that equation (\ref{bar2}) is identical to the conditions for
having a first order deformation in Altmann's construction \cite{altmann}.

Anomaly free rank distributions in the gauge theory can therefore
be built from assignments of weights $b(v)$ to all vectors $v$ in the 
(p,q) web satisfying equation (\ref{bar2}).

Consider the case when the toric CY cone has a $k-1$
dimensional branch of complex structure deformations, that is 
the toric diagram $P$ admits a Minkowski decomposition into
$k$ polytopes $P=P_1+ \ldots P_k$. 
Recall that the set
of sides of $P$ is the union of the set of sides of the summands
$P_j$: equivalently the set of vectors $v$ of the (p,q) web is split 
into $k$ disjoint subsets of vectors (let us call these sub-webs again
$P_j$) at equilibrium, that is for every sub-web $P_j$ the sum of
vectors is still zero.
As explained in Section \ref{classes},
we expect the existence of a fractional brane (rank distribution
in field theory) dual to a supergravity solution with a smoothed
deformed cone. We propose the following
conjecture for finding such rank distribution: 

\verb| |

\noindent
{\bf Deformation Fractional Branes:}
The rank distribution in the quiver gauge theory dual to the
deformation of a toric CY cone with toric diagram $P=P_1+\ldots
P_k$ is computed through a baryonic charge obtained assigning constant
weights $M_j$ to all the vectors belonging to the same sub-web $P_j$: 
$b(v)=M_j$ for $v \in P_j$, $j=1,\ldots k$.

\verb| |

Note in fact that since sub-webs are in equilibrium, equation (\ref{bar2})
is trivially satisfied, and the rank distribution will be anomaly
free.
This proposal nicely fits Altmann's rule of decomposition into a
Minkowski sum.
The rank distribution depends on $k$ arbitrary constants $M_j$, but
indeed the parameter space of deformations is $k-1$ dimensional:
recall that adding a common constant to all weights $b$ in the (p,q)
web does not change the baryonic symmetry, 
so that fractional branes are indeed
counted by the differences between the constants
$M_j$. Correspondingly in the gauge theory the rank distribution is
defined up the a common constant that can be added to all gauge groups
(regular branes).   

We have not a general proof of the above proposal, but we checked it
in many concrete examples. 
One important check that one can perform is that rank distributions
computed with the above proposal lead to a supersymmetric vacuum. 
In the case where a single deformation parameter is turned on, it is
easy to prove that with the proposed rank distributions no gauge group
develops an ADS superpotential and therefore the vacuum is expected to
be supersymmetric, see the following Section. A more refined check is
to compute the moduli space of the quiver gauge theory, probed by a single
regular brane $N=1$, and show that it is the deformed cone. We
will do this on a concrete example in Section \ref{example} along the
lines of \cite{Pinansky,susyb2}, but after having introduced the useful
tool of the $\Psi$-map. 

We point out that in general there can be different rank distributions
on the same dimer configuration (also having fixed the toric phase), 
that are dual to the same deformed geometry
with the same deformation parameters. For instance consider the
splitting of the (p,q) web in only two sub-webs at equilibrium: the
distance between their weights $b$ is an integer $M$, (number of 
fractional branes). By changing $M$ in $-M$ we find another
distribution of ranks for gauge groups, that could seem also very
different from the previous one. But in all the examples we considered 
we found
that after applying some Seiberg dualities it is possible to pass from
one distribution to the other and hence they describe the same deformed
geometry (in fact analyzing the two cascades in the far IR with a
single regular brane one can see that they reduce to the same theory).
We suggest that this is a general feature. 

Another possible ambiguity arises when the singularity is not
isolated. In this case there are parallel vectors in the (p,q) web
perpendicular to the same edge of the toric diagram. If the toric
diagram has a Minkoswki decomposition, the assignments of this
parallel vectors in the (p,q) web to the different sub-webs may be
ambiguous, and this gives rise to apparently different rank
distributions (look at Figure \ref{defspp} and at the two baryonic
charges in Figures \ref{sppcharges} a) and b)). 
However the Minkowski decomposition into polytopes is the
same and we expect a unique deformation; again we checked in the
considered examples that these ambiguities are resolved by Seiberg
dualities: also in these cases the different rank distributions are
connected by Seiberg dualities.
Therefore we conjecture that all baryonic symmetries with weights
constant on sub-webs in equilibrium compute rank distributions
dual to deformed geometries, but what matters are the absolute
distances between the weights of the sub-webs.    

\begin{figure}
\begin{center}
\includegraphics[scale=0.55]{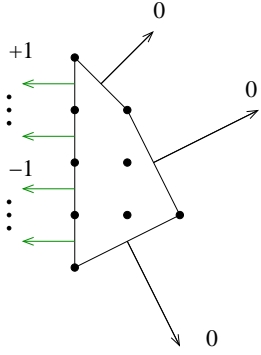}
\caption{The weights $b_i$ for $\mathcal N=2$ branes.}
\label{n2}
\end{center}
\end{figure}

Let us now turn to the case of $\mathcal N=2$ fractional branes.
Consider a toric diagram with one edge $E$ having $k-1$ integer
internal points (see Figure \ref{n2} for the case $k=4$). 
This corresponds to a surface of singularities of
$ \mathbb{C}^2/ \mathbb{Z}_k$ type. In the (p,q) web there are $k$
vectors $w_1, \ldots w_k$ perpendicular to the edge $E$.
Our proposal is:

\verb| |

{\bf $\mathcal N=2$ fractional branes:}
The rank distributions in the quiver gauge theory corresponding to 
$\mathcal N=2$ fractional branes are computed by baryonic symmetries
obtained assigning weights $b(w_j)=b_j$, $j=1, \ldots k$ for the $k$ 
vectors $w_j$ perpendicular to $E$ with the constraint
$b_1+ \ldots b_k=0$, and $b(v)=0$ for all
other vectors $v$ in the (p,q) web.

\verb| |

Again recall that we can add a common constant to all $b_i$ of these
configurations and have still the same baryonic symmetry, and hence the
same rank distribution.
Note that this choice obviously satisfies equations (\ref{bar2})
for baryonic charges since we have imposed that the sum of weights
$b_j$ is zero. Moreover there is a space of $k-1$
independent $\mathcal N=2$ fractional branes as expected for 
$ \mathbb{C}^2/ \mathbb{Z}_k$ singularities. Only parameters $a_i$
associated with integer points along the edge $E$ or with its vertices 
are different from zero.  
We will check this assignment on concrete examples in the following
subsection.   

To conclude we suggest that different assignments of baryonic charges,
not associated with splittings of (p,q) web or with edges with
integer points, correspond in general to SB fractional branes.

\section{Examples and further observations}
\label{check}

\begin{figure}
\begin{center}
\includegraphics[scale=0.40]{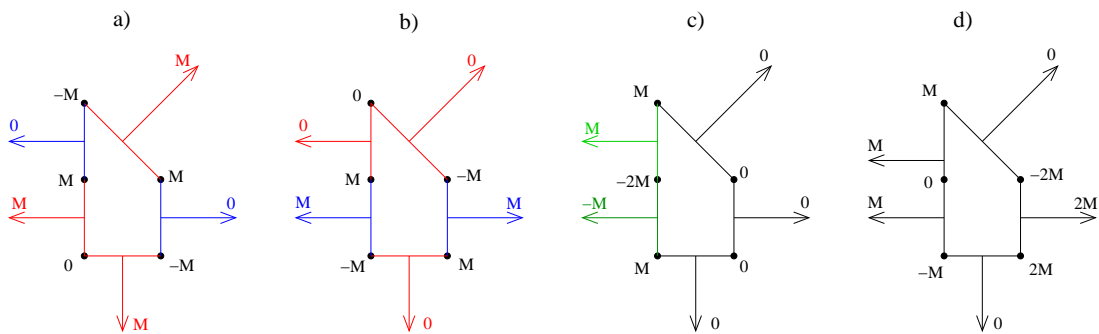}
\caption{The weights for baryonic charges in SPP theory.
a) Deformation brane; rank distribution: $(N,N+M,N)$. b) Deformation
brane; rank distribution: $(N,N,N+M)$. c) $\mathcal N=2$ brane; rank
distribution $(N+M,N,N)$. d) SB brane; rank distribution:
$(N+M,N,N+2M)$.}
\label{sppcharges}
\end{center}
\end{figure}

Let us start with the example of the SPP. It has $d=5$ and hence has a two
dimensional space of fractional branes, but only a one dimensional
space of complex deformations. It is known in the literature 
that the rank distribution $(N,N+M,N)$ for gauge groups (1,2,3)
reported in Figure \ref{spp2}, corresponds to a
deformation brane. In fact it is easy to see from Figures \ref{spp1}
and \ref{spp2} that this corresponds to
the choice of baryonic charges $(a_1,a_2,a_3,a_4,a_5)$ =
$(-M,M,-M,M,0)$ or equivalently $(b_1,b_2,b_3,b_4,b_5)$
=$(M,0,M,0,M)$. These charges are reported in Figure \ref{sppcharges}
a), from which it is evident that the weights $b$ for sub (p,q) webs
in equilibrium are constant. The gauge theory undergoes a cascade of
Seiberg dualities that reduce the number of regular branes $N
\rightarrow N-M$. In the IR, if there are no more regular branes (that
is $M$ divides $N$) we can put $N=0$ and so only a single
confining $SU(M)$ gauge group survives (the second one).
This is the case $M>0$.

If instead $M$ is negative, in the IR we
have to put $N=|M|$ and we get two $SU(|M|)$ gauge groups (groups 1
and 3, whereas groups 2 disappears) and a superpotential term
$W_0=-X_{11}X_{13}X_{31}$. By
performing a Seiberg duality\footnote{Gauge group 3 has $N_f=N_c=|M|$
  in the IR, so that it is not possible to perform a Seiberg duality;
  yet the moduli space of vacua is quantum modified: $W=W_0+X(det\,
  N-B\bar B - \Lambda ^{2M})$, where $X$ is a Lagrangian multiplier,
  $B$, $\bar B$ the baryons and $N=X_{13}X_{31}$ the meson matrix. Along the
  baryonic branch: $X=0$, $B=i \Lambda^M \xi$, $\bar B= i\Lambda^M /
  \xi$ the gauge group condensates, and the superpotential becomes
  $W=-X_{11}N$. These massive fields can be integrated out in the IR. 
  From a diagrammatic point of view this is formally
  equivalent to perform a Seiberg duality. This is the same reason
  that allows to delete some gauge groups at the end of the cascade
  when all regular branes disappear.} with respect to face 3 (the
square) we come back to a single isolated gauge group.

There is another equivalent distribution of ranks corresponding to the
deformation brane: $(b_1,b_2,b_3,b_4,b_5)$ = $(0,M,0,0,M)$; this is
reported in Figure \ref{sppcharges} b): it corresponds again
to a subdivision into two subwebs in equilibrium. The rank
distribution corresponding to this baryonic symmetry is
$(N,N,N+M)$. Since the exchange of gauge groups 2 and 3 is a symmetry
of this theory (look the dimer from upside down) it is easy to see
that this gauge theory is equivalent to the previous one. Again M can
be also negative.

Since SPP is not an isolated singularity, there is also an $\mathcal
N=2$ fractional brane: it is known that the
associated rank distribution for the gauge groups is $(N+M,N,N)$,
and this corresponds to the choice of baryonic charge
$(b_1,b_2,b_3,b_4,b_5)$ = $(0,0,0,M,-M)$, in agreement with our proposal.

In Figure \ref{sppcharges} d) we report a choice of baryonic charges 
that gives the rank
distribution: $(N+M,N,N+2M)$. For this type of fractional brane a
supersymmetric vacuum is not present:
group 3 has $N_f<N_c$ (with N=0) and generates a non perturbative ADS
superpotential leading to runaway behavior. 
The corresponding baryonic symmetry $b_i:$ $(0,2M,0,M,M)$ 
is not associated with a splitting of
the (p,q) web in subwebs at equilibrium. Note that this configuration
can be obtained as a linear combination of the two deformation branes
in Figure \ref{sppcharges} a) and \ref{sppcharges} b) up to a global
constant for all $b_i$: generic superpositions of fractional
deformation branes that do not satisfy the criterion in Section
\ref{matching} lead to SB. This fact was already noted in \cite{susyb1}. 

\begin{figure}
\begin{center}
\includegraphics[scale=0.55]{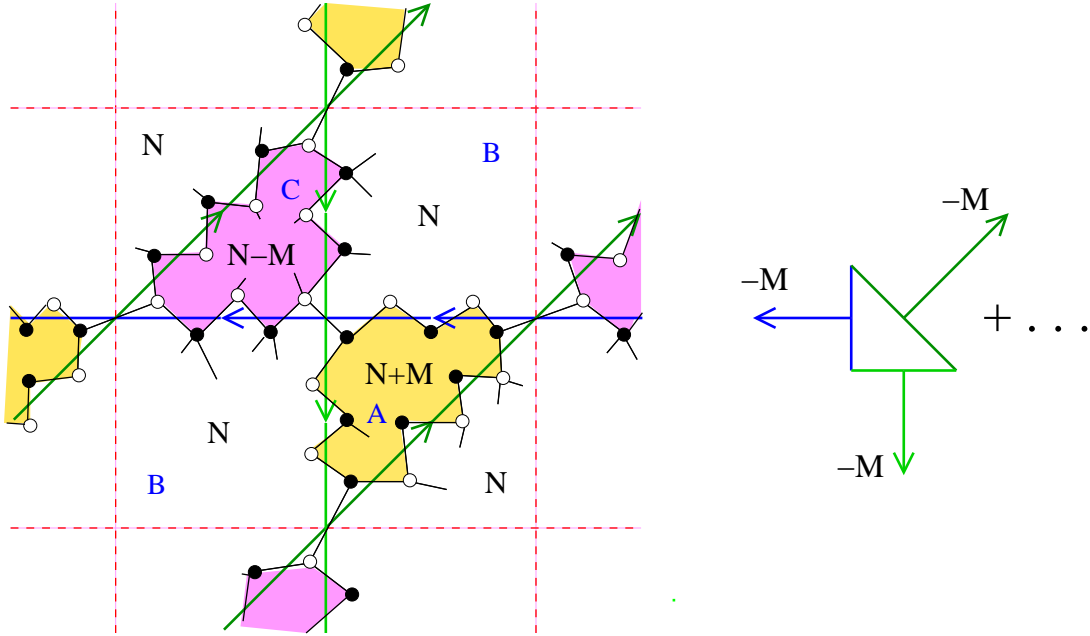}
\caption{The rank distribution for a gauge theory dual to a toric
  geometry obtained by the Minkowski sum of a triangle and another
  polygon. More generally if there is only one deformation pa\-ra\-me\-ter the
  gauge groups can be only $SU(N)$, $SU(N+M)$ and $SU(N-M)$.}
\label{due}
\end{center}
\end{figure}

Consider now the general case when the (p,q) web is splitted into two
different sub-webs $P_1$ and $P_2$ at equilibrium (if further splittings are
allowed we turn on a single deformation parameter). 
According to our proposal the deformational fractional brane is
computed by a baryonic symmetry with weights $b$ of the type:
$b(v)=-M$ if $v \in P_1$, $b(v)=0$ if $v \in P_2$. Look at
Figure \ref{due}, where for simplicity $P_1$ is a triangle.  

To reconstruct the dimer we have to draw the zig-zag paths
corresponding to vectors of $P_1$ and $P_2$ as in \cite{rhombi}, with
the suitable consistency conditions. It is interesting to note that
in the complete dimer for $P_1+P_2$, the zig-zag paths corresponding
to $P_1$ (or $P_2$) satisfy separately the consistency conditions: for
example in Figure \ref{due}, if we isolate the three zig-zag paths
associated with vectors of $P_1$ (lines in light green, dark green and
blue) we see that they divide the fundamental cell of the torus
(delimited by the red dashed lines) into three regions: a face (in
magenta) where the zig-zag paths turn clockwise, another face (in
yellow) where the zig-zag paths turn anticlockwise, and a third face
with an even number of sides (where zig-zag paths are not
oriented). This is just the way in which the Fast Inverse Algorithm
reconstructs the theory associated with the triangle $P_1$, the
$\mathcal N=4$ SYM: the non oriented face is the gauge group,
clockwise oriented face is the white vertex, and anticlockwise
oriented face is the black vertex. 

This is a general feature of dimers dual to a toric diagram that can
be splitted in $P_1+P_2$ and has been recently noted also in \cite{uranga},
where it was explained in the context of mirror
symmetry and using ideas from geometric 
transition.\footnote{Developing ideas from \cite{multiflux,susyb1}, in 
  the same paper \cite{uranga} the sub-webs
  splitting at the level of dimers was also used to show that it is
  possible to  ``deform'' the theory for $P_1+P_2$ to, say, the theory of
  $P_1$. To obtain this, one has to choose mesonic vevs to move the
  regular branes in the deformed space. In our paper instead we
  do not give vevs to mesonic operators, but, analogously to the
  Klebanov-Strassler case, we consider the cascades on the baryonic
  branches. Therefore in the IR, for deformation branes, when all
  regular branes have disappeared, we typically
  find confining gauge groups with no matter left. To be clear we say
  also that in this paper we consider cascades where only the number
  of regular branes is decreased: to have multiple cascades as in
  \cite{multiflux} one has to turn on mesonic vevs.} 

For a generic polytope $P_1$ we will call $C$ ($A$) the
regions along which zig-zag paths of $P_1$ turn clockwise (anticlockwise) and
$B$ the non-oriented regions. In general $A$, $B$, $C$ are unions of
contractible regions in the torus $T^2$; a region of type $A$ (or $C$)
is rounded only by region(s) of type $B$.  

In Figure \ref{due} we have drawn only the zig-zag paths associated
with $P_1$; their intersections correspond to links in the dimer that
separate regions of type $B$ and have baryonic charge zero. But
there are other links: we have drawn also all the links of the 
dimer along the zig-zag paths of $P_1$: they correspond to an
intersection of a zig-zag paths of $P_1$ with one of $P_2$.
Note that inside the regions $C$ ($A$) there are only
white (black) vertices belonging to the zig-zag paths of $P_1$ because
zig-zag paths turn clockwise (anticlockwise) around white (black)
nodes; there could be however ``more interior'' vertices of different
colors inside $C$ and $A$, not belonging to the zig-zag paths of $P_1$.

The links in the dimer correspond to intersections of two zig-zag
paths: if the zig-zag paths are associated with (p,q) web vectors of
the same sub-web $P_i$, then the baryonic charge of the link is zero
because of equation (\ref{carica}), otherwise the charge is $M$ or $-M$.
Therefore the only charged links under the baryonic symmetry are those
separating regions of type $A$ from regions of type $B$ and regions
$C$ from $B$. 
Let us assign number of colors $N$ to all faces in regions $B$; then
using the rules and the conventions explained in Section
\ref{matching} and in Figure \ref{charge} we can deduce that all faces in
regions $A$ will have number of colors $N+M$ and regions $B$ number of
colors $N-M$.

So our proposed baryonic symmetry gives rise only to gauge groups
$SU(N)$, $SU(N+M)$ or $SU(N-M)$ (one of these could be absent as we
will see). 
If we suppose the existence of a cascade,
in the IR (when $M$ divides $N$), we can put $N=M$. At this step all
regular branes have disappeared and we have only gauge groups $SU(M)$
in regions $B$ and $SU(2M)$ in regions $A$. 
These gauge groups cannot develop ADS terms: since $N_f$ and $N_c$ are
both multiples of $M$ the only problem would be an $SU(2M)$ gauge
group with $M$ flavors. But region $A$ is rounded by region $B$ and so
an $SU(2M)$ gauge group is rounded at least by four faces (it is at
least a square) with colors
at least $SU(M)$, therefore it has $N_f\geq 2M$. Obviously ADS terms
cannot appear in previous steps of the cascade: if we add regular
branes (in multiples of $M$) then for every gauge group $N_f$
increases faster than $N_c$.

Since no ADS term is generated we expect in these cases that
supersymmetry is not spontaneously broken. In concrete examples we
found that, also when regular branes have disappeared, it is possible
to continue to perform some Seiberg duality (gauge group condensation)
until we are left with only isolated confining gauge groups. 

If for example there are no faces in regions $C$ (no $SU(N-M)$ gauge
groups) the analysis is easier: in the IR we can put N=0, so that we
are left only with $SU(M)$ gauge groups in regions $A$. Again there are
no ADS superpotential terms ($N_f=0$ or $N_f\geq M$). Note that
superpotential terms due to vertices inside regions $A$ typically
allow to perform gauge condensations until only confining groups and no
massless chiral matter survives in the IR.    

\begin{figure}
\begin{center}
\includegraphics[scale=0.74]{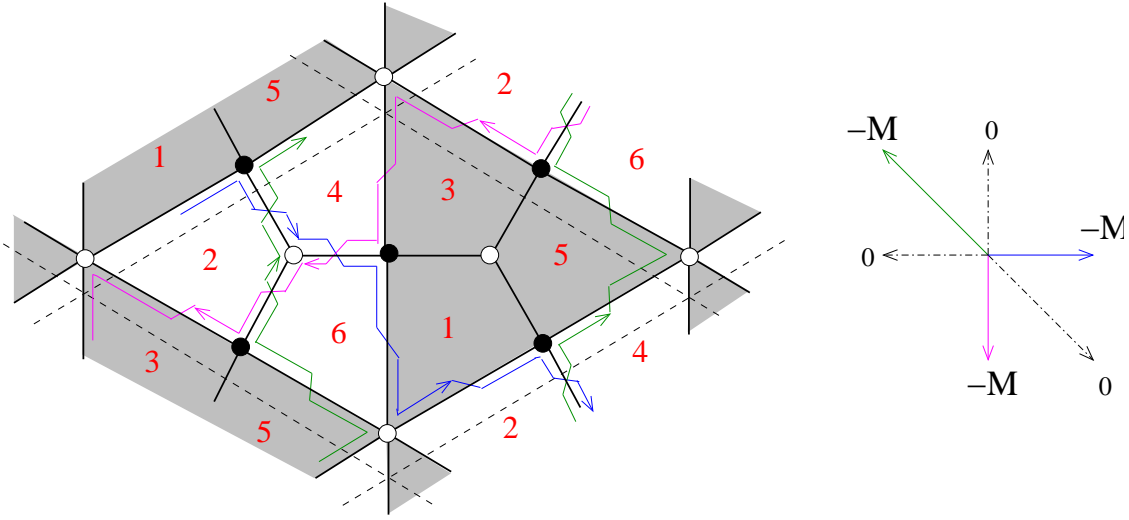}
\caption{The rank distribution for deformation of $dP_3$,
corresponding to Figure \ref{defdp3} a). The shadowed gauge groups
1,3 and 5 are $SU(N+M)$. The other groups, 2,4,and 6, are $SU(N)$.}
\label{dp3branch1}
\end{center}
\end{figure}

We can check these ideas in the known case of $dP_3$
\cite{multiflux,susyb1}. For the first deformation branch in Figure
\ref{defdp3} a) we draw the rank distribution in Figure
\ref{dp3branch1}, where we show the three zig-zag paths corresponding
to the edges of one of the triangles in the Minkowski sum of the toric
diagram. Note that in this case the region of type $C$ contains only a
white vertex and no faces, so that there are only $SU(N)$ gauge groups
(faces 2,4,6 in regions $B$) and $SU(N+M)$ gauge groups (faces 1,3,5
in region $A$). In the IR we can put $N=0$ and we have the three
$SU(M)$ gauge groups 1,3,5 with the corresponding superpotential
term. Performing a Seiberg duality (gauge group condensation on the baryonic
branch) with respect to one of them and integrating out massive fields
we have two confining gauge groups in the IR.

Other cases with a single deformation parameter $P=P_1+P_2$ that have
only $SU(N)$ and $SU(N+M)$ gauge groups are the cases where $P_1$ is a
segment: the (p,q) web of $P_1$ is a pair of opposite vectors, see
Figure \ref{parallel1} below. 

\begin{figure}
\begin{center}
\includegraphics[scale=0.6]{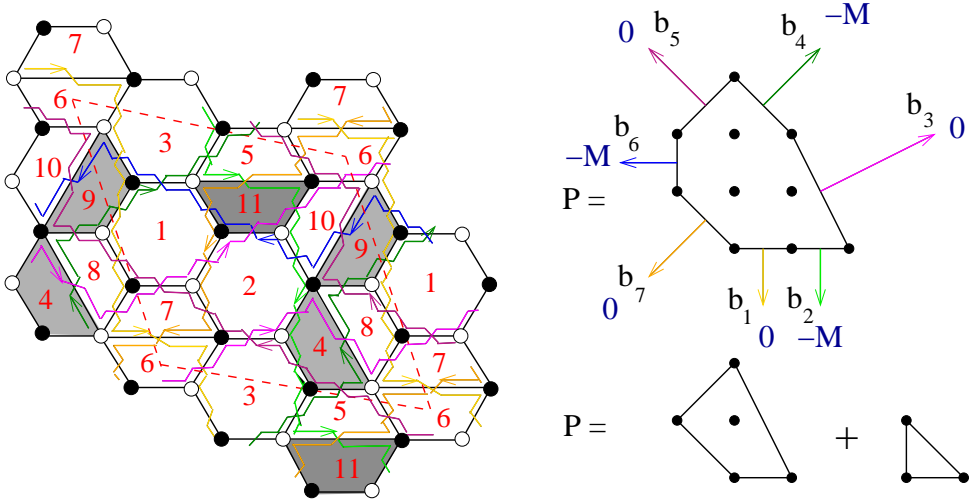} 
\caption{Dimer configuration dual to the toric diagram 
 $P=Y^{2,1} + \mathbb C^3$.}
\label{y21t}
\end{center}
\end{figure}

We give a concrete example of the general case in Figure \ref{y21t}:
we consider a toric diagram $P$ obtained by summing the toric
diagram of $P_1 \equiv \mathbb C^3$ (a triangle) and of $P_2 \equiv
Y^{2,1}$. We constructed
a minimal toric phase, reported in Figure \ref{y21t}, with the Fast
Inverse Algorithm \cite{rhombi}. There are $11$ gauge groups labelled in
red; the red dashed lines delimit the fundamental cell. You can see
that assigning weights $b_i$: $(0,-M,0,-M,0,-M,0)$ to the zig-zag
paths we obtain the rank distribution for the fractional deformation
brane reported in Figure \ref{y21t}
with $SU(N+M)$ for faces 9,4 (type $A$ regions); $SU(N-M)$ for face 11
(type $C$ region), and $SU(N)$ for the remaining gauge groups.

Note that since this is not an isolated singularity we have the
ambiguity described it the previous Section: we can assign also
weights $b_i$: $(-M,0,0,-M,0,-M,0)$ and obtain an equivalent rank
distribution with only $SU(N-M)$ and $SU(N)$ gauge groups. We report
the possible rank distributions for this theory in the following
table:
\[
\begin{small}
\begin{array}{c|ccccccccccc}
(b_1,b_2,b_3,b_4,b_5,b_6,b_7) & 1 & 2 & 3 & 4 & 5 & 6 & 7 & 8 & 9 & 10
  & 11\\[0.5em] \hline 
(0,-M,0,-M,0,-M,0) & N & N & N & N+M & N & N & N & N & N+M & N
  & N-M \\
(-M,0,0,-M,0,-M,0) & N & N & N & N & N-M & N-M & N & N-M & N & N-M
  & N-M\\
(0,M,0,M,0,M,0) & N & N & N & N-M & N & N & N & N & N-M & N
  & N+M \\
(M,0,0,M,0,M,0) & N & N & N & N & N+M & N+M & N & N+M & N & N+M
  & N+M
\end{array}
\end{small}
\]
where we have added in the last two lines also the possibilities of
exchanging $M$ with $- M$, (in the following: $M>0$).
    
The four distributions above may seem at first glance to be different;
indeed we checked for all of them the existence of a cascade of
Seiberg dualities: the dimers come back to themselves up to a permutation
of groups with $N \rightarrow N-M$. At the end of the respective
cascades we find (if $M$ divides $N$) that gauge groups condensate
until there remain always three isolated confining
gauge groups. If instead we consider the case with one regular brane
remaining  in the IR (like in Section \ref{example}) the four gauge
theories reduce to the same theory in the IR, so that they are dual to
the same deformed geometry. More generally it is possible to find
Seiberg dualities that send each configuration in the previous table
to one another. 

\begin{figure}
\begin{center}
\includegraphics[scale=0.6]{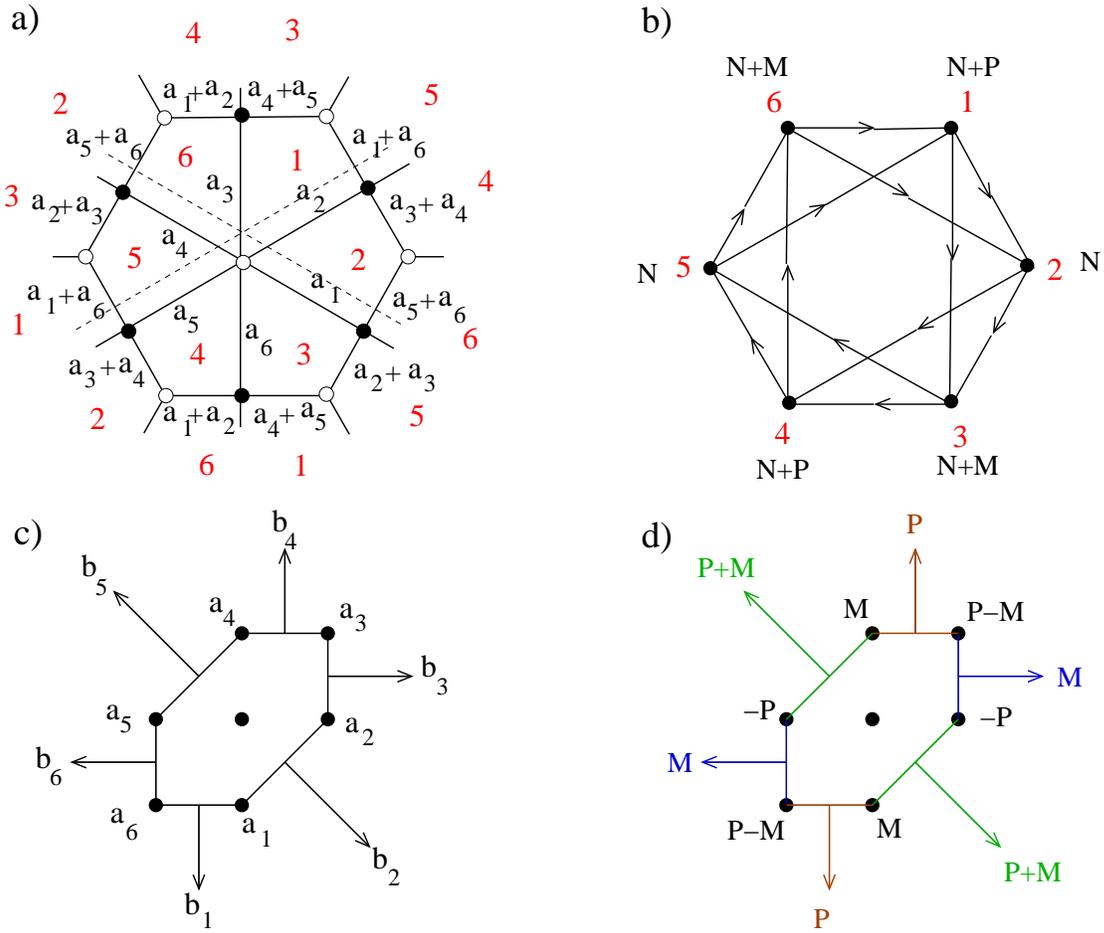}
\caption{The deformation of $dP_3$ corresponding to Figure
  \ref{defdp3}b). a) Dimer configuration and distribution of charges. 
  b) Quiver and rank assignment. c) Toric Diagram. d)
  Values of $a_i$ and $b_i$ for the baryonic charge corresponding to
  this rank distribution.}
\label{dp3branch2}
\end{center}
\end{figure}

Deformations with more parameters are in general more difficult to
treat. In Figure \ref{dp3branch2} we report the rank distribution for
the deformation of the cone over $dP_3$ corresponding to Figure
\ref{defdp3} b). This is a two parameters branch and correspondingly
we have two integers, $P$ and $M$ parametrizing the weights for
the zig-zag paths: $b_i=$ $(P,P+M,M,P,P+M,M,P)$. Note that our
proposal for the baryonic charge of deformation branes reproduces the
known results in the literature, see Figure \ref{dp3branch2} b).

\begin{figure}
\begin{center}
\includegraphics[scale=0.50]{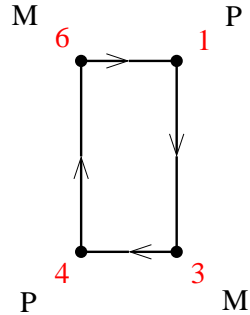}
\caption{The quiver gauge theory for $dP_3$ in the IR with only
  fractional branes, corresponding to Figure \ref{dp3branch2} b).}
\label{dp3quiver}
\end{center}
\end{figure}

In the IR, when all regular branes have disappeared ($N=0$), there
remain the four gauge groups $1,3,4,6$ with ranks respectively
$P,M,P,M$, see Figure \ref{dp3quiver}, and the tree level
superpotential term:
\begin{equation}
W_0=-\mathrm{tr} \left( X_{61} X_{13} X_{34} X_{46} \right)
\end{equation}  
differently from the case with a single deformation parameter, we see
that ADS superpotential terms may appear. In our example, if $P>M>0$,
groups 1 and 4 have $N_f<N_c$ and so the superpotential becomes,
\begin{equation}
W= -\mathrm{tr} \left( M_{63} M_{36} \right) + 
c \left( \frac{1}{\mathrm{det} M_{63}} \right)^{\frac{1}{P-M}}+
d \left( \frac{1}{\mathrm{det} M_{36}} \right)^{\frac{1}{P-M}}
\end{equation}
where $M_{63}$ and $M_{36}$ are the $M \times M$ mesonic matrices of
groups $1$ and $4$ respectively. It is easy to see that F-term and
D-term equations can be satisfied (choose the meson matrices
proportional to the identity). Therefore there exists a supersymmetric
vacuum for this theory.

\begin{figure}
\begin{minipage}[t]{0.48\linewidth}
\centering
\includegraphics[scale=0.67]{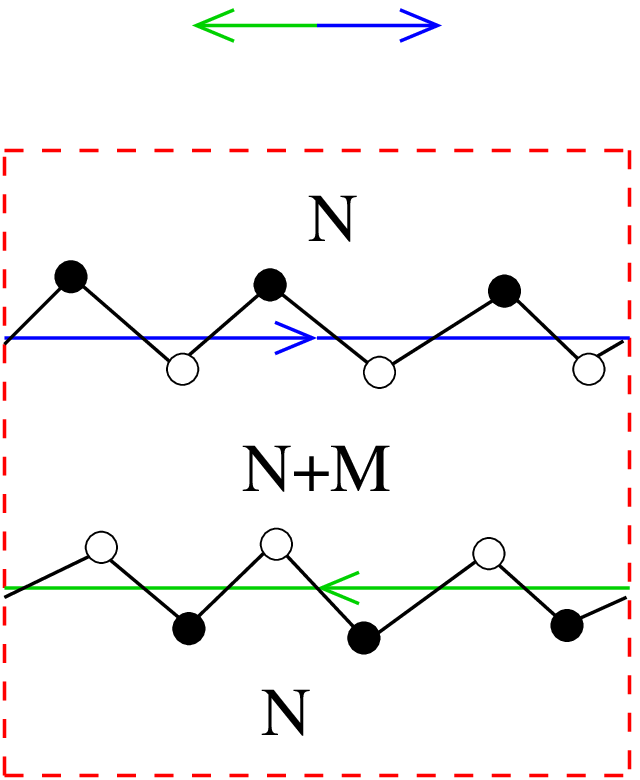}
\caption{Rank distribution for a fractional deformation brane obtained
  by lifting a subweb made up of two opposite vectors.}
\label{parallel1}
\end{minipage}%
~~~~~~\begin{minipage}[t]{0.48\linewidth}
\centering
\includegraphics[scale=0.67]{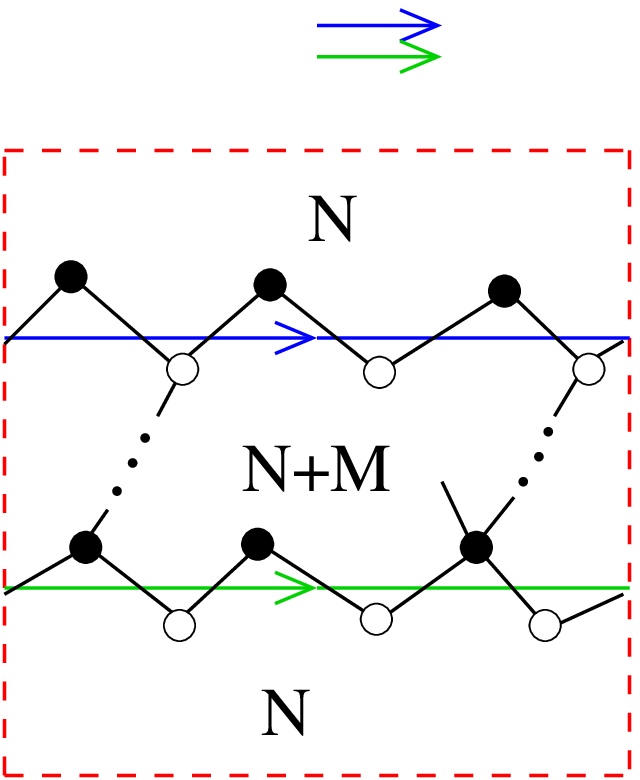}
\caption{Rank distribution for an $\mathcal N=2$ fractional brane
  obtained by giving opposite weights to two (p,q) web vectors 
  perpendicular to the same edge of the toric diagram.}
\label{parallel2}
\end{minipage}%
\end{figure}

Let us now make some further comments on the $\mathcal N = 2$
fractional branes. In Figures \ref{parallel1}, \ref{parallel2}  
we compare the case of a
fractional deformation brane obtained by lifting two opposite vectors
in the (p,q) web with the same weights (Figure \ref{parallel1}), and
the case of an $\mathcal N=2$ fractional brane obtained by giving
opposite weights to two parallel vectors in the (p,q) web
(perpendicular to the same edge of the toric diagram), and weight zero
to all other vectors. In both cases the fundamental cell of the torus
(delimited by the red dashed lines in the Figures) is divided in two
strips where faces have ranks $N$ and $N+M$ respectively. We have
drawn only the white and black vertices belonging to the two zig zag
paths we are considering: for the deformation brane inside the strip
with $N+M$ ranks there fall the white vertices: one would need an even
number of links to pass from one white vertex of the first zig-zag
path to a white vertex of the second zig-zag path; among these
configurations there are also those consisting of only isolated
groups. Instead for the $\mathcal N=2$ fractional brane, inside the strip
with $N+M$ ranks there fall the white vertices of the first zig-zag
path and the black ones of the second zig-zag path: now an odd number
of links is required to pass from one set of vertices to the
other. Consider for example the $\mathcal N = 2$ fractional brane
obtained with $b_1=M$ and $b_2=-M$ in Figure \ref{y21t}, and zero to the
remaining $b_i$. If we do not insert regular branes ($N=0$), we have a
closed loop of $SU(M)$ gauge groups connected by chiral fields 
(faces 6,5,4,8,9,10) but
with no superpotential term. If we give vev to all but one chiral
fields the theory reduces to a single gauge group with an adjoint
multiplet.  

The analysis just performed fits the observation already done in
\cite{susyb1} that $\mathcal N=2$ fractional branes
correspond to rank distributions where faces with $N+M$ rank form
parallel strips on the torus: this is due to the fact that we give 
weights only to parallel vectors in the (p,q) web, that correspond to
parallel not intersecting zig-zag paths. 

\begin{figure}
\begin{center}
\includegraphics[scale=0.58]{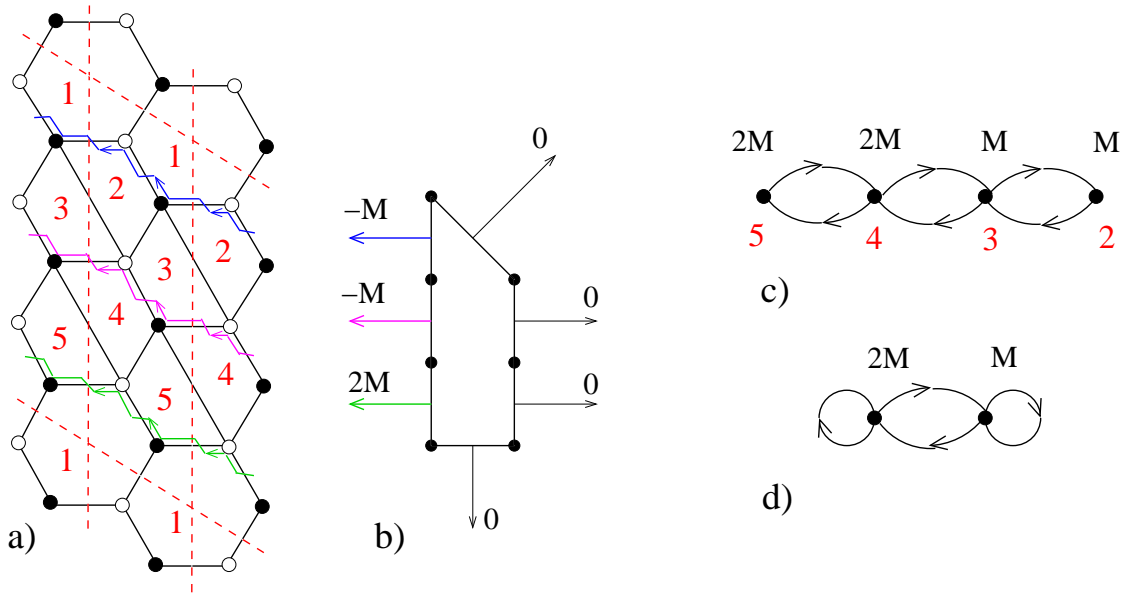}
\caption{$\mathcal N=2$ fractional branes in the theory for the cone
  over $L^{2,3;3,2}$. a) Dimer configuration. b) Toric diagram and
  weights for the baryonic symmetry. c) Quiver gauge theory
  (with $N=0$). d) Quiver after giving vevs to $X_{23}$ and $X_{45}$.}
\label{l2332}
\end{center}
\end{figure}

In Figure \ref{l2332} we give a more complicated example of an
$\mathcal N=2$ fractional brane: the theory is $L^{2,3;3,2}$ whose
dimer and toric diagram are reported in Figure \ref{l2332} a) and b)
respectively. We draw in the dimer only the three zig-zag paths that
correspond to the vectors perpendicular to the fourth edge of the
toric diagram. You can check that giving weights that sum up to zero
only to these three vectors you obtain theories that have regions of
the moduli space with an accidental $\mathcal N=2$ supersymmetry. For
instance if we give weights $(-M,-M,2M)$ we obtain the rank
distribution $(N,N+M,N+M,N+2M,N+2M)$ for the five gauge groups. If
there are no regular branes $N=0$, the quiver reduces to that drawn in
Figure \ref{l2332} c) with the superpotential:
\begin{equation}
W=X_{23}X_{34}X_{43}X_{32}-X_{45}X_{54}X_{43}X_{34}
\label{sup}
\end{equation}
Giving vev to, for instance, $X_{23}$ and $X_{45}$ and integrating out
massive fields the quiver reduces to that reported in Figure
\ref{l2332} d): there are two gauge groups with adjoints and an
hypermultiplet of $\mathcal N=2$. Also the superpotential (\ref{sup})
reduces to that of an $\mathcal N=2$ theory with matter.

\section{Comments on the $\Psi$-map}
\label{psimap}

In this Section we introduce the correspondence between the chiral
ring of mesonic operators in the (superconformal) field theory and 
the semigroup of integer points in $C^*$, the dual cone of the fan $C$. 
This correspondence has already been studied in the literature 
\cite{exceptional,kru}, see also \cite{bhms}; here we simply
translate these results in the language of charges and zig-zag paths. 
Then we provide a direct proof in field theory that the $\Psi$-map 
of a mesonic operator is an affine function, and make further useful
comments. 

The idea of the correspondence is simple: the moduli space of a 
superconformal quiver gauge theory
(we can restrict to the abelian case $N=1$ with a single 
regular brane) is the toric CY cone 
where the D3-brane can move in the string theory set up; 
this toric cone is described by a convex rational
polyhedral cone in $\mathbb R^3$, the fan $C$.
Mesonic operators (closed oriented loops) in field theory can be 
considered as well defined functions on the toric cone: the value
of the function in every point of the moduli space is the vev of 
the mesonic operator in that vacuum.
Obviously such functions are the same for F-term equivalent operators.
In toric geometry the ring of algebraic functions on the toric cone 
is in one to one correspondence with the semigroup of integer points 
in $C^*$, the dual cone of the fan $C$ (generated by inward pointing 
normals to the faces of $C$).
\begin{equation}
C^* = \{ x \in \mathbb R^3 \, | \, (x,y) \geq 0 \quad \forall y \in C \}
\end{equation}
We may then expect a one to one correspondence between the chiral 
ring of mesonic operators, equivalent up to F-terms, and the 
integer points in $C^*$. This is indeed the case, and, as we will see,
the three integer numbers that are associated to mesons are (related 
to) the charges of the meson under the three global $U(1)$ isometries 
of the geometry, that are the flavor symmetries and the R-symmetry 
in field theory. A precise mapping can be obtained using the 
$\Psi$-map, introduced in \cite{exceptional}.  

Let us consider an oriented link $X$ in the periodic quiver 
(or in the dimer). As explained in Sections \ref{distribution} 
and \ref{matching}, one can parametrize the charges of the link 
$X$ with a formal\footnote{Here we are ignoring the restriction 
(\ref{sum}) and its analogue for R-symmetries.} expression:
\begin{equation}
\Psi(X)=\sum_{i=1}^d c_i a_i
\label{psi}
\end{equation}
where the coefficients $c_i$ are integer numbers (0 or 1
for a single link) 
that can be easily computed using one of the two algorithms 
reviewed in Section \ref{distribution}.
The $\Psi$-map associates to every oriented link $X$ a function 
$\Psi(X)$ defined on the vertices (and integer points on the 
boundary) of the toric diagram $P$ of the dual geometry. This function
evaluated at vertex $i$ is defined as: $\Psi_i(X)=c_i$.
In the following we shall use the same expression $\Psi(X)$ for this
function and for the expression in (\ref{psi})\footnote{Formally we
  are substituting the symbols $D_i$ of the divisors associated with
  the $i$-th vertex \cite{exceptional} with the parameters $a_i$ 
  for charges. Therefore
  we do not make the quotient with respect to affine functions in the
  plane of the toric diagram.}.

As in \cite{exceptional} we linearly extend the $\Psi$ map to the
group of one-chains $L$ in the periodic quiver, the free group generated
by the oriented links $X_j$ in the quiver with integer coefficients:
\begin{equation}
L=\sum_j d_j X_j \qquad d_j \in \mathbb Z
\label{cycles}
\end{equation}  
In particular the $\Psi$ map of a path in the quiver is obtained by
summing the trial charges of fields in the path (or by subtracting the
charges for links that are not oriented in the same direction as the path).

Note that this definition is equivalent to that given in
\cite{exceptional}: for a path $L$, $\Psi_i(L)$ is the number of
intersections (weighted with +1 or -1 according to orientation) of $L$
with the perfect matching in the dimer corresponding to the $i$-th
vertex of the toric diagram. This intersection number is just the
coefficient $c_i$ of $a_i$ in the expression of the $\Psi$-map for
$L$: $\Psi(L)=\sum_i c_i a_i$,
according to the algorithm in \cite{aZequiv} for distributing charges.

Note that, for non isolated singularities, it is possible to extend the
function $\Psi(X)$ to all the integer points along the boundary of the
toric diagram (this can be done unambiguously using the algorithm in
\cite{proc} for distributing charges, in particular using the
conventions in (\ref{relation}) and (\ref{carica})).

Let us fix a system of coordinates for the fan $C$, such that the
generators have third coordinate equal to one: the integer points
along the boundary of the toric diagram are: $V_i=(x_i,y_i,1)$,
$i=1,\ldots d$. This defines the linear functions $x$ and $y$ in the
plane of the toric diagram.
As already observed in \cite{exceptional,kru}, the $\Psi$-map of a
closed loop $L$ in the periodic quiver is an affine function:
\begin{meson}
\label{meson}
If $L$ is a closed loop of chiral fields (oriented or not), then the
$\Psi$-map for $L$ is:
\begin{equation}
\Psi(L)=\sum_{i=1}^d \left( n x_i + m y_i + c \right) a_i
\label{psiL}
\end{equation}
where $(n,m,c)$ are integer numbers and $(n,m)$ are the homotopy
numbers of the loop $L$ on the torus $T^2$ of the periodic quiver or dimer. 
\end{meson} 
We give here a simple proof of this statement. From the definition
(\ref{psi}) it is obvious that the $\Psi$-map of any one cycle is a
homogeneous degree-one polynomial in the $d$ variables $a_i$. 
We may then restrict to the $d-1$ global charges imposing the
constraint (\ref{sum}) and prove that for a closed loop:
$\Psi(L)=\sum_{i=1}^d \left( n x_i + m y_i \right) a_i$. Then the general
case can only differ from this expression for an integer constant $c$
times the sum of all $a_i$. 
Now if (\ref{sum}) holds we can parametrize the global charges as in
Section \ref{matching} using the parameters $b_i$ associated with
zig-zag paths and related to $a_i$ through (\ref{relation}). The
charge of a generic link can be computed as in (\ref{carica}): as
shown in Figure \ref{charge} we have to add the weight $b$ if the topological 
intersection between L and the zig-zag path is $+1$ or subtract the
weight $b$ if the intersection is $-1$. Let us call $w\equiv (n,m)$ the
homotopy numbers of the loop L on the torus $T^2$; the homotopy
numbers of the zig-zag paths are given by the vectors $v_i$ of the
(p,q) web, $v_i=(p_i,q_i)$. 
The topological intersection between $L$ and the $i$-th zig-zag path
is $\mathrm{det}(w,v_i)$, and summing them with the weights $b_i$ we
get the global charge of $L$:
\begin{equation}
\Psi(L)= \sum_{i=1}^d \mathrm{det} \left( w,v_i \right) \, b_i  
\qquad \mathrm{if}
\quad \sum_{i=1}^d a_i=0
\label{ps}
\end{equation}
Note that this is true also if the links in $L$ are not all oriented 
in the same direction.  
If $\tilde{V}_i \equiv (x_i,y_i)$ are the coordinate vectors of the
integer points along the perimeter of the toric diagram, the vectors
$v_i$ can be obtained by a $90^o$ rotation of the edges of the toric
diagram:
\begin{equation}
v_i= R \left( \tilde{V}_i - \tilde{V}_{i-1} \right)
\label{vi}
\end{equation}
where $R$ is the rotation matrix:
\begin{equation}
R = \left(
\begin{array}{ll}
0 & 1 \\
-1 & 0
\end{array}
\right)
\end{equation}  
Indexes $i$ will be understood to be periodic of period $d$ and in our
conventions are displaced as in Figure \ref{weights}. Substituting
(\ref{vi}) into (\ref{ps}) we compute:
\begin{eqnarray}
\Psi(L) & = & \sum_i \mathrm{det} \left( w, \, R \tilde{V}_i \, b_i - R
\tilde{V}_{i-1} \, b_i \right) \nonumber \\
 &  = & \sum_i \mathrm{det} \left(w, R \tilde{V}_i \, (b_i-b_{i+1})
\right) \nonumber \\
 &  = & - \sum_i \mathrm{det} \left( w, R \tilde{V}_i \right) a_i
\label{case}\\
 &  = & \sum_{i} (n x_i + m y_i)\, a_i \qquad \qquad \quad \mathrm{if}
\quad \sum_{i=1}^d a_i=0 
\end{eqnarray}
where in the third equality we have used the relation
(\ref{relation}). This concludes the proof. Note as a particular case
that equality (\ref{case}) together with (\ref{bar}) also shows that the 
baryonic charge of a closed loop is zero, as claimed in Section
\ref{fractional}.

Mesonic operators in field theory are the trace of chiral fields along
an oriented closed loop $L$. In the oriented case the coefficients
$d_i$ in (\ref{cycles}) are all non negative, and hence the
coefficients $c_i$ in $\Psi(L)=\sum_i c_i a_i$ are all non
negative. Because of theorem (\ref{meson}) the $c_i$ are the scalar
products:
\begin{equation}
c_i=\left( (n,m,c),(x_i,y_i,1) \right), \qquad \qquad c_i \geq 0 \,\,\,
\mathrm{for}\,\, \mathrm{oriented}\,\, \mathrm{loops.}
\label{scalar}
\end{equation}
We see therefore that the vector $(n,m,c)$ lies in $C^*$ for
mesons. Hence the explicit mapping from mesons to
integer points in $C^*$ is just given by the $\Psi$-map (\ref{psiL}):
$L \leftrightarrow (n,m,c)$. 

Note that this correspondence is well defined under changes of
coordinates: the coefficients $c_i$ of the $\Psi$-map (\ref{psi}) do
not depend on the choices of coordinates, hence if we perform a
translation or an $SL(2,\mathbb Z)$ transformation of the toric
diagram, the point $(n,m,c)$ transforms as a point of the dual lattice
$C^*$ to keep the scalar product $c_i$ in (\ref{scalar}) constant.

Since F-term equivalent mesons have the same charges they are mapped
to the same point $(n,m,c)$. Conversely if mesons are mapped to the
same point $(n,m,c)$ it means that they have the same homotopy numbers
in the torus and the same ``length'' (i.e. the same R-charge in the
language of \cite{exceptional}, this is obvious since the $\Psi$-map
is just a parametrization of the R-charge). Using Lemma 5.3.1 (that
makes use of the hypotheses of consistency of the tiling) in
\cite{exceptional} we conclude that such mesons are F-term equivalent.
Moreover we suppose that the $\Psi$-map is surjective on $C^*$;
the work of \cite{exceptional} also suggests that the correspondence
is one to one.

As a consequence mesons in the chiral ring and integer points of the additive
semi-group $C^*$ also satisfy the same algebraic relations: if two
linear combinations with positive integer coefficients of integer
vectors in $C^*$ are equal, then the composition of the corresponding
mesons are still closed oriented loops with the same image under the
$\Psi$-map, since it is linear, and are therefore $F$-term equivalent.
Recall that in toric geometry to every independent generator of the semi-group
$C^*$ is associated a complex variable $z_j$ and that linear relations
between generators become equations that express the toric cone as a
(non complete) intersection in the space of $z_j$. Conversely in field
theory the moduli space (in the case with a single regular brane
$N=1$) can be computed through algebraic and F-term relations between mesonic
operators (for concrete examples see the following Section or the work
of \cite{Pinansky}). We have just seen through the $\Psi$-map that the two
kinds of computations always agree in the toric superconformal case:
a consistent dimer configuration built according to the rules of
the Fast Inverse Algorithm \cite{rhombi} has a moduli space of vacua
that always reproduces the dual toric geometry. Therefore we point out
that \emph{the $\Psi$-map theory can work as an argument to show directly that
  the Fast Inverse Algorithm is correct}\footnote{For a beautiful
  justification of the Fast Inverse Algorithm in the context of mirror
symmetry see \cite{mirror}. A rigorous proof of the Fast Forward Algorithm
based on perfect matchings can be found in \cite{forward}.}. 
In fact the proofs of
disposition of charges, of Theorem \ref{meson}, and of Lemma 5.3.1 in
\cite{exceptional} are based only on the assumption that the dimer is
built according to the rules of the Fast Inverse Algorithm: zig-zag
paths must be closed non intersecting loops, they are in one to one
correspondence with the legs of the (p,q) web and must be drawn on a
torus  with the homotopy numbers (p,q) of the corresponding leg; links in the
dimer are in one to one correspondence with intersections of two
zig-zag paths; zig-zag paths turn clockwise around white vertices and
anticlockwise around black vertices... 

In the following Section we shall see how the $\Psi$-map works on a
concrete example; moreover the $\Psi$-map greatly simplifies the
problem of finding the mesons corresponding to generators of $C^*$.
Then we shall consider the deformed moduli space in presence of
fractional branes. To conclude we note that the vector $(0,0,1)$
always belongs to $C^*$ since the generators $V_i$ are
$(x_i,y_i,1)$. This vector will play an important role in the study of
deformations. In the superconformal case the mesons that are mapped to
this vector are the superpotential terms (recall that their R-charge
is 2, and hence their $\Psi$ map is $\sum_i a_i$); they are all F-term
equivalent as it is easy to prove directly or using the fact that they
are mapped to the same vector under $\Psi$-map.

\section{Computing the deformed geometry: a detailed example}
\label{example}

Up to now we have checked in examples that our prescription in Section
\ref{matching} for fractional deformation branes leads to a
supersymmetric vacuum: in the IR if no regular branes survive after the
cascade we find isolated confining gauge groups. Indeed one should
check that also the deformed geometry is reproduced by the gauge
theory. We will do that on a specific example, along the lines of
\cite{Pinansky}, by considering the case
of a single regular D3 brane in the IR, plus the fractional
branes. The moduli space of the gauge theory, deformed by the presence
of ADS terms, describes the geometry probed by the D3 brane and
therefore should match with the geometry of the deformed cone that can
be computed through Altmann's algorithm \cite{altmann}.

\begin{figure}
\begin{center}
\includegraphics[scale=0.6]{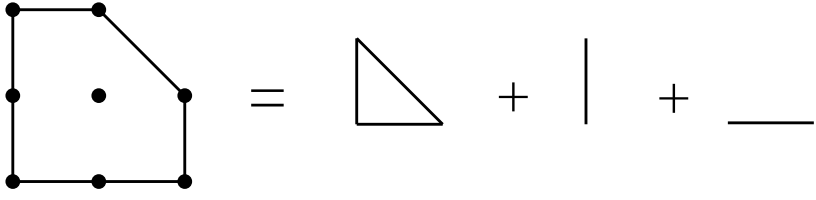}
\caption{The toric diagram for the cone over $PdP_4$ and its Minkowski
  decomposition.}
\label{pdp4}
\end{center}
\end{figure}

The example we have chosen is the 
$PdP_4$ theory, whose toric diagram has vertices:
\begin{equation}
(0,0,1) \quad (2,0,1) \quad (2,1,1) \quad (1,2,1) \quad (0,2,1)
\label{coneC}
\end{equation}
and admits a Minkowski decomposition into a triangle and two segments,
see Figure \ref{pdp4}, corresponding to a two dimensional deformation
branch. 

\begin{figure}
\begin{center}
\includegraphics[scale=0.6]{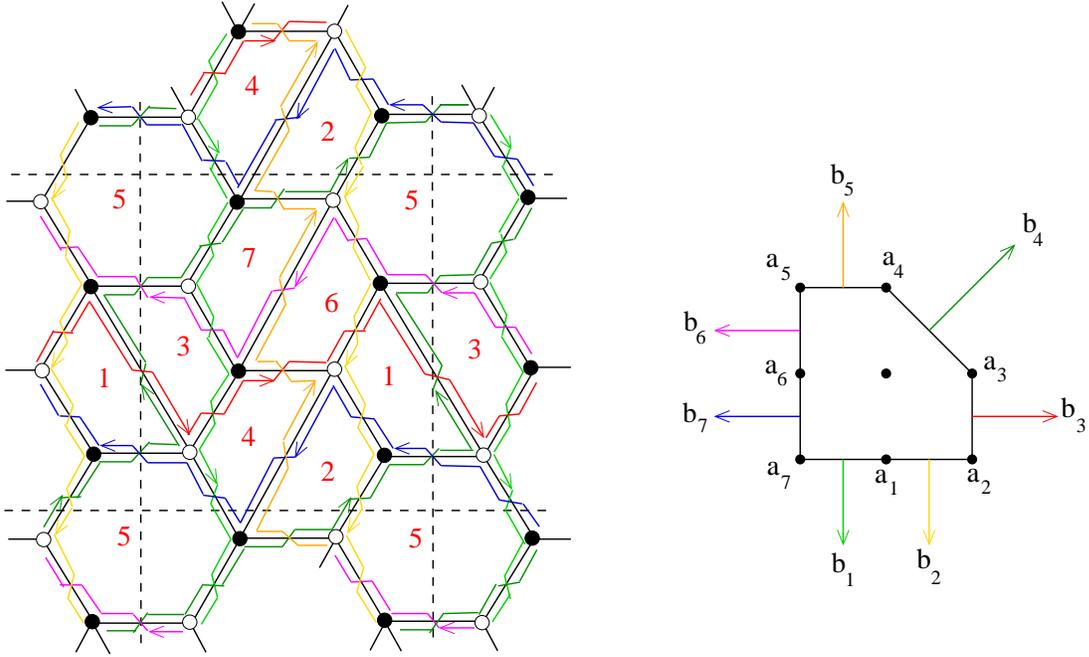}
\caption{The dimer configuration for $PdP_4$. We report on the right 
also the toric diagram for $PdP_4$ with the legs of the (p,q) web colored as
the corresponding zig-zag paths in the dimer.}
\label{triqdimer}
\end{center}
\end{figure}

\begin{figure}
\begin{center}
\includegraphics[scale=0.6]{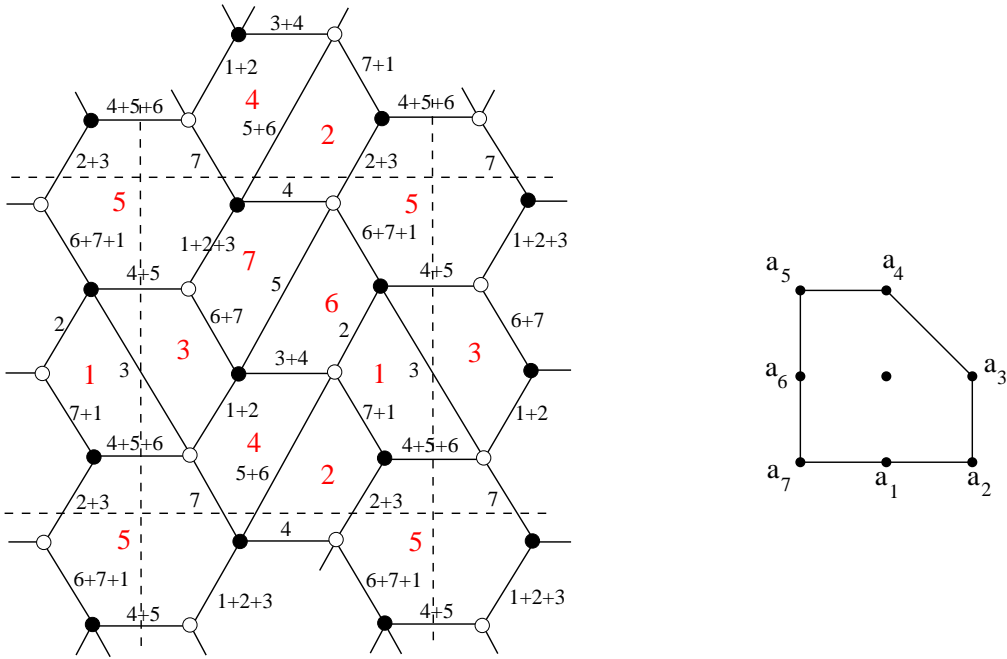}
\caption{The distribution of charges $a_i$ for $PdP_4$. In the dimer a
black number $i$ near a link stands for the charge $a_i$.}
\label{triqcharges}
\end{center}
\end{figure}

The dimer can be easily reconstructed through the Fast Inverse
Algorithm or by looking in the literature: we report it in
Figure \ref{triqdimer}, where we draw also the zig-zag paths and their
correspondence with vectors in the (p,q) web. This is a minimal toric
phase. There are $7$ gauge groups
labelled in red: we have chosen this labelling so as to reproduce the
quiver in Figure 15 of \cite{susyb1}. The fundamental cell is
delimited by the dashed black lines in the dimer.
From the zig-zag paths it is easy to reconstruct the charge
distribution for links in the dimer, as explained in Sections
\ref{distribution} 
and \ref{matching}; we draw it in Figure \ref{triqcharges}.


Let's start with the superconformal case. First of
all we have to compute the cone $C^*$: the inward pointing
perpendiculars to the faces of (\ref{coneC}) are:
\begin{equation}
z_1 \rightarrow (0,1,0) \quad
z_2 \rightarrow (-1,0,2) \quad
z_3 \rightarrow (-1,-1,3) \quad
z_4 \rightarrow (0,-1,2) \quad
z_5 \rightarrow (1,0,0) 
\label{gen1}
\end{equation}  
they are primitive integer vectors.
These vectors generate $C^*$ over $\mathbb R^+$, but do not generate
the lattice cone of integer points in $C^*$ over positive integer
numbers. In our case we have to add another generator:
\begin{equation}
t \rightarrow (0,0,1)
\label{gen2}
\end{equation} 
We assign complex variables $z_1$, $z_2$, $z_3$, $z_4$, $z_5$, $t$ to
the six generators as in (\ref{gen1}) and (\ref{gen2}).

Linear relationships satisfied by the generating vectors (linear
combination with positive integer coefficients equal to another combination
with positive integer coefficients) are translated into complex equations
for the corresponding variables. A minimal set of relations is in our
case: 
\begin{equation}
\begin{array}{lll}
z_1 z_3 = z_2 t & z_2 z_4= z_3 t & z_3 z_5=z_4 t\\
z_2 z_5 =t^2 & z_1 z_4=t^2 & 
\end{array}
\label{singular}
\end{equation}
where for example the last equation translates the linear relation:
$(0,1,0)$ $+$ $(0,-1,2)$ $=$ $2(0,0,1)$. 
The relations in (\ref{singular}) define the toric cone over $PdP_4$ 
as a non complete intersection in $\mathbb C^6$. Note in fact that
there are 6 variables and 5 relations, but the cone has complex
dimension 3: 2 relations depend on the others at generic points where
$z_i$ and $t$ are not zero. In our example all generators lie on a
plane (this is not true in general) and so it is simpler to find the
relations.  

To see that the moduli space of gauge theory (with $N=1$) reproduces
the singular geometry (\ref{singular}), we have to compute the F-term
relations in the chiral ring of mesonic operators. 

\begin{figure}
\begin{center}
\includegraphics[width=0.6\textwidth, height=0.6\textwidth]{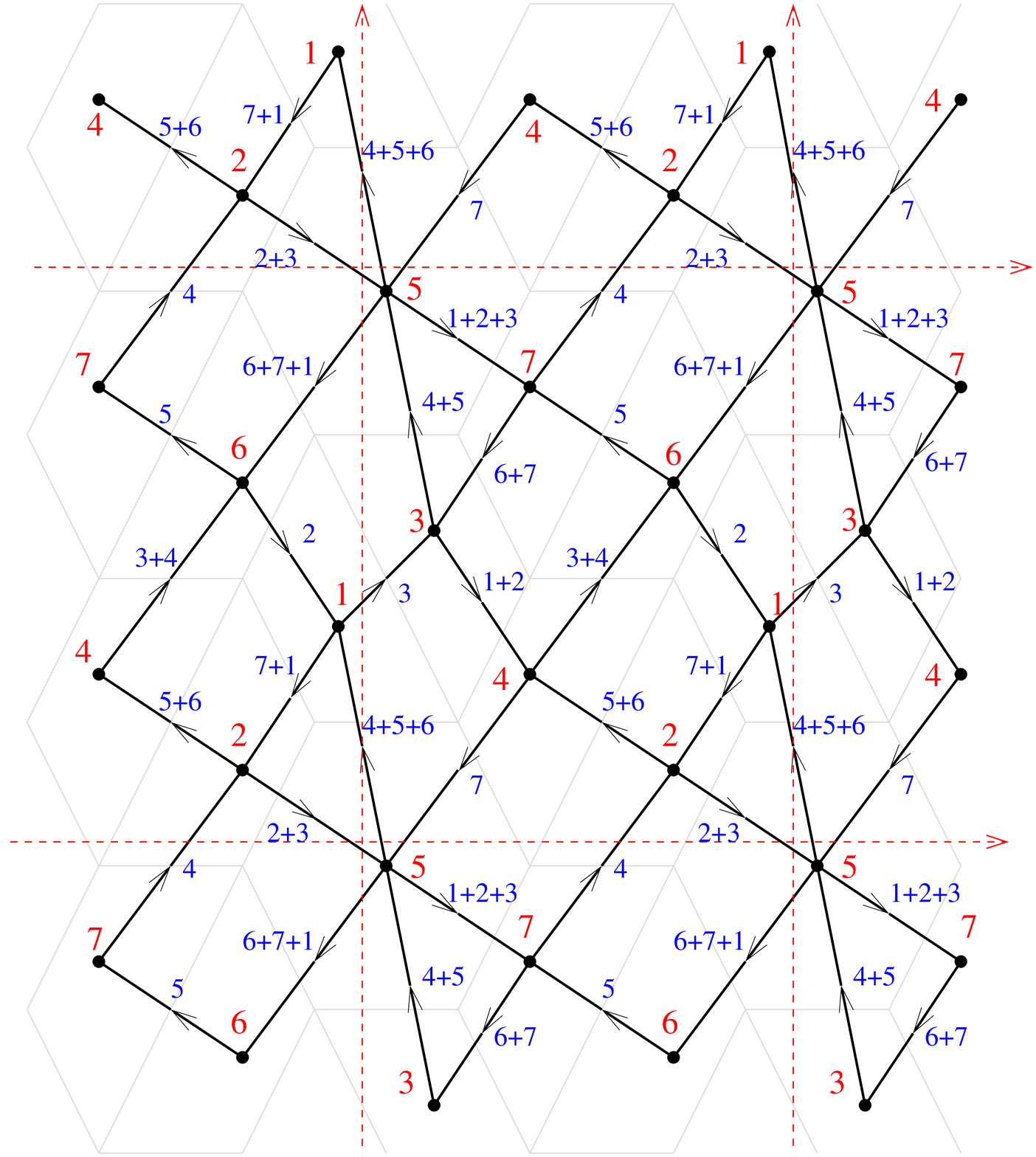}
\caption{The periodic quiver for $PdP_4$. Red numbers label gauge
  groups, blue numbers label charges $a_i$.}
\label{quiver}
\end{center}
\end{figure}

To find mesons that
correspond to the generating vectors of $C^*$, one can use the
$\Psi$-map theory and Theorem \ref{meson} in Section
\ref{psimap}. For example consider the first vector $z_1
\rightarrow(0,1,0)$. We know that a meson $Z_1$ mapped to this vector by the
$\Psi$-map has homotopy numbers given by the first two entries $(0,1)$, and
$\Psi$-map equal to the linear function $y$ in the plane of the toric
diagram, that is $\Psi(Z_1)=\sum_i y_i a_i= a_3+a_6+2a_4+2a_5$ (look
at Figure \ref{triqdimer} for the disposition of $a_i$ in the plane of
the toric diagram). So one can look for a path in the dimer or in the
periodic quiver (drawn in Figure \ref{quiver}) with homotopy numbers
$(0,1)$ of minimal length (adding loops would add factors of $\sum_i
a_i$ to the $\Psi$-map). For example you can check that the meson
$Z_1=X_{46}X_{67}X_{72}X_{24}$ has all these features and therefore it
is $\Psi$-mapped to $(0,1,0)$, as one can check explicitly.
Any other meson with the same $\Psi$-map is F-term equivalent. 

\begin{figure}
\begin{center}
\includegraphics[scale=0.6]{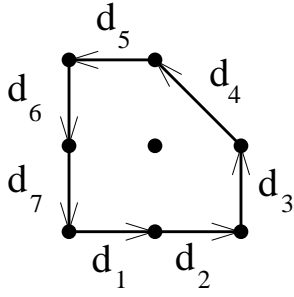}
\caption{The vectors $d_j$ for the toric diagram $PdP_4$.}
\label{tdef}
\end{center}
\end{figure}

Continuing in this way, one can find the following representatives for
mesons corresponding to generating vectors:
\begin{equation}
\begin{array}{l|l|l|l|l}
    & (n,m,c) & n x+m y+c & \Psi-\textrm{map} & \textrm{meson}\\ \hline
z_1 & (0,1,0) & y & a_3+2a_4+2a_5+a_6 & Z_1=X_{46}X_{67}X_{72}X_{24} \\
z_2 & (-1,0,2)&  -x+2 & a_1+a_4+2a_5+2a_6+2a_7 & Z_2=X_{35}X_{56}X_{67}X_{73} \\
z_3 & (-1,-1,3)& -x-y+3 & 2a_1+a_2+a_5+2a_6+3a_7 & Z_3=X_{45}X_{56}X_{67}X_{73}X_{34} \\
z_4 & (0,-1,2) & 2-y & 2a_1+2a_2+a_3+a_6+2a_7 & Z_4=X_{45}X_{57}X_{73}X_{34} \\
z_5 & (1,0,0) & x & a_1+2a_2+2a_3+a_4 & Z_5=X_{57}X_{72}X_{25} \\
t &   (0,0,1) & 1 & a_1+a_2+a_3+a_4+a_5+a_6+a_7 & T=X_{12}X_{25}X_{51}
\label{mesoni}
\end{array}
\end{equation}
We understand the traces in writing mesons since when $N=1$ the 
fields are complex numbers.
The superpotential $W_0$ can be read directly from the dimer:
\begin{eqnarray}
W_0& = & X_{56}X_{67}X_{72}X_{25}+X_{73}X_{35}X_{57}+X_{46}X_{61}X_{12}X_{24}
+X_{13}X_{34}X_{45}X_{51} \nonumber \\
 & & - X_{57}X_{72}X_{24}X_{45}-X_{13}X_{35}X_{56}X_{61}
-X_{46}X_{67}X_{73}X_{34}
 -X_{51}X_{12}X_{25}
\end{eqnarray}
Using the F-term equations $\partial W_0 / \partial X$ one can show
that the mesons in (\ref{mesoni}) satisfy the same relations
(\ref{singular}): 
\begin{equation}
\begin{array}{lll}
Z_1 Z_3 = Z_2 T & Z_2 Z_4= Z_3 T & Z_3 Z_5=Z_4 T\\
Z_2 Z_5 =T^2 & Z_1 Z_4=T^2 & 
\end{array}
\end{equation}
In Section \ref{psimap} we gave arguments based on the $\Psi$-map to show that 
this matching is true in general in the superconformal case. 

Let us now consider the deformed case; we will start to compute 
the deformed geometry corresponding to the Minkowski decomposition in
Figure \ref{pdp4}.

\subsection{The deformed geometry}
We will follow Altmann's work \cite{altmann} (for a brief 
account of the algorithm see \cite{Pinansky}). Note however that
we are extrapolating the algorithm to the case of a not isolated
singularity.

We will label with $d_j$, $j=1,\ldots,7$, the vectors along the perimeter 
of the toric diagram and with $t_j$ their weights, see Figure \ref{tdef}:
\begin{equation}
\begin{array}{llll}
d_1=d_2=(1,0) & d_3=(0,1) & d_4=(-1,1) & d_5=(-1,0) \\
d_6=d_7=(0,-1) & & & 
\end{array}
\end{equation}

A solution for the $t_i$ to the deformation conditions is:
\begin{equation}
\begin{array}{l}
t_1=t_7=t_4=t\\
t_2=t_5=t+S_1\\
t_3=t_6=t+S_2
\end{array}
\label{parametrization}
\end{equation}
corresponding to the Minkowski decomposition in Figure \ref{pdp4}.
It will be easy to see that equivalent parametrizations of the $t_i$
(obtained by exchanging $t_1$ with $t_2$ and $t_6$ with $t_7$) will
not give rise to ambiguities in the final equations.

\begin{figure}
\begin{center}
\includegraphics[scale=0.4]{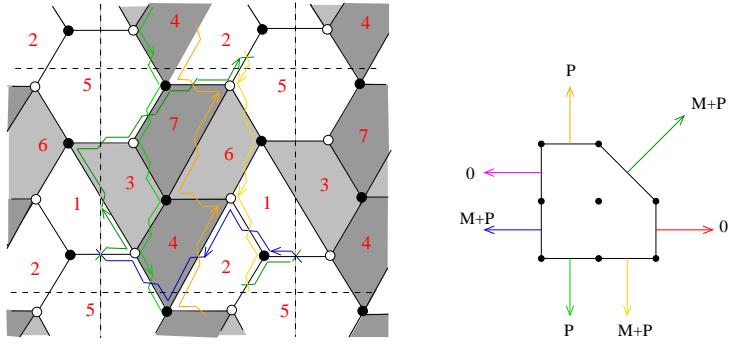}
\caption{The rank distribution corresponding to $b_1=b_5=P$,
  $b_2=b_4=b_7=M+P$, $b_3=b_6=0$. The gauge groups are $SU(N)$ for
  faces 1, 2, 5; $SU(N+M)$ for faces 4, 7; $SU(N+M+P)$ for faces 6, 3.}
\label{distribution1}
\end{center}
\end{figure}

Now the algorithm is the following: write every generating vector of 
$C^*$, with the exception of $(0,0,1)$, in the form $(c^i,\eta_0 (c^i))$,
with $c^i$ given by the first two components and $\eta_0 (c^i)$ the third
component. Find a point $a(c^i)$ along the perimeter of the toric diagram
satisfying $a(c^i) \cdot c^i+ \eta_0 (c^i)=0$. Then find a path
representation for the point $a(c^i)= \lambda^i_j d_j$ and compute for
every $i$ the vector $\eta(c^i)=(-\lambda^i_1 (d_1 \cdot c^i), \ldots,
-\lambda^i_7 (d_7 \cdot c^i))$. We report the results in the following
table: 
\begin{equation}
\begin{array}{llclll}
 & \hspace{\stretch{1}} c^i \hspace{\stretch{1.5}} & \eta_0(c^i) &
 a(c^i) &\hspace{\stretch{1}} \lambda^i
  \hspace{\stretch{1}} & \hspace{\stretch{1}}\eta (c^i)\hspace{\stretch{1}} \\
z_1 & (0,1) & 0 & (0,0) & (0,0,0,0,0,0,0) & (0,0,0,0,0,0,0) \\
z_2 & (-1,0) & 2 & (2,0) & (1,1,0,0,0,0,0) & (1,1,0,0,0,0,0) \\
z_3 & (-1,-1) & 3 & (2,1) & (1,1,1,0,0,0,0) & (1,1,1,0,0,0,0) \\
z_4 & (0,-1) & 2 & (1,2) & (1,1,1,1,0,0,0) & (0,0,1,1,0,0,0) \\
z_5 & (1,0) & 0 & (0,2) & (1,1,1,1,1,0,0) & (-1,-1,0,1,1,0,0)
\end{array}
\end{equation}
Now every equation in (\ref{singular}) is replaced in the following
way:
\begin{equation}
t^a \prod_i z_i ^{p_i}=\prod_i z_i^{q_i} \quad \rightarrow \quad
\prod_i t_i^{\left(\sum_j q_j \eta(c^j)-p_j \eta(c^j) \right)_i}
 z_i^{p_i}=\prod_i z_i^{q_i}
\end{equation}
and one can show that the degree is conserved: $a= \sum_i
\left( \sum_j q_j \eta(c^j)-p_j \eta(c^j) \right)_i$. Substituting the $t_i$
with the parametrization (\ref{parametrization}), we finally find the
equations in the deformed case:
\begin{equation}
\begin{array}{lll}
z_1 z_3 = z_2 (t+S_2) & z_2 z_4= z_3 t & z_3 z_5=z_4 (t+S_1)\\
z_2 z_5 =t (t+S_1) & z_1 z_4=t (t+S_2) & 
\label{def}
\end{array}
\end{equation}
which still define a three dimensional complex geometry. 

\subsection{The moduli space of the gauge theory}

First of all we have to compute the rank distribution in the gauge
theory due to the fractional branes using our proposal in Section
\ref{matching}. As already said there are different ways to do that,
exchanging the weights $b_1$ and $b_2$ or $b_6$ and $b_7$, but they
all lead to a supersymmetric vacuum\footnote{For toric (pseudo) del Pezzos
surfaces another method to find rank distributions for fractional
deformation branes was used in \cite{multiflux,susyb1}: basically
for simple toric diagrams with one internal point the number of legs
in the (p,q) web is equal to the number of gauge groups in the dual
theory (double area), and one can define a correspondence between
them, and hence a rank distribution can be assigned fitting Altmann's
rule. We have seen that for $dP_3$ this method provides the
same results than that proposed in Section \ref{matching}. Instead, as
already noted in \cite{susyb1}, for
$PdP_4$ the algorithm in \cite{multiflux,susyb1} do not give the
right fractional deformation branes for all the correct choices of
weights of (p,q) web legs. So for $PdP_4$ we have to use the general
algorithm in Section \ref{matching}.}. 

\begin{figure}
\begin{center}
\includegraphics[scale=0.4]{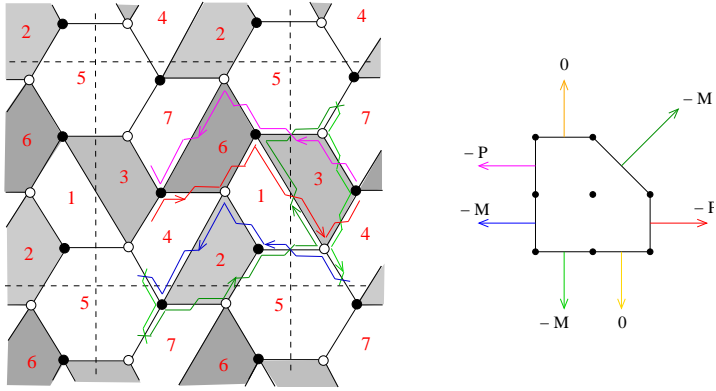}
\caption{The rank distribution corresponding to $b_3=b_6=-P$,
  $b_1=b_4=b_7=-M$, $b_2=b_5=0$. The gauge groups are $SU(N)$ for
  faces 1, 4, 7, 5; $SU(N+M)$ for face 2; $SU(N+P)$ for face 6;
  $SU(N+P-M)$ for face 3.}
\label{distribution2}
\end{center}
\end{figure}

In Figure \ref{distribution1} for example we report the choice:
$(P,M+P,0,M+P,P,0,M+P)$ for the $b_i$. Using the charge distribution
in Figure \ref{triqcharges} it is easy to see that the ranks are:
$(N,N,N+M+P,N+M,N,N+M+P,N+M)$. In the IR if $N=0$ we have a
configuration equal to that of Figure \ref{dp3quiver}, and hence we 
expect a supersymmetric vacuum.

In Figure \ref{distribution2} we draw the rank distribution for
another possible choice of $b_i$: $(-M,0,-P,-M,0,-P,-M)$ which leads
to the ranks: $(N,N+M,N+P-M,N,N$, $N+P,N)$. We will consider the case
$P>M>0$. In the IR if $N=0$ there survive only the three isolated
groups 2,3, and 6. Therefore this configuration is easier and we will
study the moduli space in this case.

We have to consider a single $N=1$ regular brane. In the IR the non
abelian gauge groups are 2,3 and 6 with ranks $SU(M+1)$,
$SU(P-M+1)$ and $SU(P+1)$ and they all have $N_f<N_c$, developing ADS
superpotential terms.
We replace the chiral fields connected to these groups with the mesons
$A,B,C$ of gauge groups $2,3,6$ respectively: 
\begin{equation}
\begin{array}{l}
A=\left( 
\begin{array}{ll}
A_{14} & A_{15} \\
A_{74} & A_{75}
\end{array}
\right) \equiv
\left(
\begin{array}{ll}
X_{12}X_{24} & X_{12}X_{25} \\
X_{72}X_{24} & X_{72}X_{25} 
\end{array}
\right) \\[0.3em]
B=\left( 
\begin{array}{ll}
B_{14} & B_{15} \\
B_{74} & B_{75}
\end{array}
\right) \equiv
\left(
\begin{array}{ll}
X_{13}X_{34} & X_{13}X_{35} \\
X_{73}X_{34} & X_{73}X_{35} 
\end{array}
\right) \\[0.3em]
C=\left( 
\begin{array}{ll}
C_{41} & C_{47} \\
C_{51} & C_{57}
\end{array}
\right) \equiv
\left(
\begin{array}{ll}
X_{46}X_{61} & X_{46}X_{67} \\
X_{56}X_{61} & X_{56}X_{67} 
\end{array}
\right) 
\end{array}
\end{equation} 
and the superpotential is:
\begin{eqnarray}
W & = &
  A_{75}C_{57}+B_{75}X_{57}+C_{41}A_{14}+B_{14}X_{45}X_{51}\nonumber \\
  &   & -A_{74}X_{45}X_{57}-C_{51}B_{15}-B_{74}C_{47}-A_{15}X_{51}
  \nonumber \\
  &   & +\alpha \log (\det A)+\beta \log (\det B)+\gamma \log (\det C)
\label{super}
\end{eqnarray}
where we have rewritten $W_0$ through mesons and have added the three
ADS terms using the glueballs $\alpha$, $\beta$, $\gamma$ that will be
matched to the two deformation parameters $S_1$ and $S_2$, similarly
as in \cite{Pinansky}.
The mesons $Z_i$ and $T$ can be rewritten as:
\begin{equation}
\begin{array}{lll}
Z_1=C_{47} A_{74} & Z_2=B_{75}C_{57} & Z_3=X_{45}C_{57}B_{74} \\
Z_4=X_{45}X_{57}B_{74} & Z_5=X_{57}A_{75} & T=A_{15}X_{51}
\end{array}
\label{Zm}
\end{equation}
where again we do not write traces because, using mesons $A,B,C$, all
fields are abelian.
We will check that the deformed equations (\ref{def}) are
satisfied, focusing on generic points where all the mesons are
different from zero. It is then easy to write the F-term equations from
(\ref{super}) and invert them to express some fields in function of the
others. For example we found:
\begin{equation}
\begin{array}{lll}
A_{14}=\displaystyle
\frac{A_{74}(A_{74}X_{45}X_{57}-\alpha)}{A_{75}X_{51}} &
A_{15}=\displaystyle \frac{A_{74}X_{45}X_{57}}{X_{51}} &
B_{75}=\displaystyle A_{74}X_{45} \\[1em]
B_{74}=\displaystyle \frac{A_{74}X_{45}X_{57}+\beta}{C_{47}} &
B_{15}=\displaystyle \frac{A_{74}C_{47}}{X_{51}} &
B_{14}=\displaystyle \frac{A_{74}X_{57}}{X_{51}}\\[1em]
C_{51}=\displaystyle \frac{X_{51}(A_{74}X_{45}X_{57}+\beta)}{A_{74}C_{47}} &
C_{57}=\displaystyle \frac{A_{74}X_{45}X_{57}-\alpha}{A_{75}} &
C_{41}=\displaystyle \frac{A_{75}X_{51}}{A_{74}} 
\end{array}
\label{fterm}
\end{equation}
and moreover F-term equations imply the relation:
\begin{equation}
\gamma=\alpha+\beta
\label{gamma}
\end{equation}
so that indeed there are only two independent parameters.

Substituting into the explicit expressions for mesons (\ref{Zm}) the
results from F-term conditions (\ref{fterm}) and (\ref{gamma}), it is
easy to prove that the mesons $Z_i$ satisfy relations analogous to
(\ref{def}): 
\begin{equation}
\begin{array}{lll}
Z_1 Z_3 = Z_2 (T+\beta) & Z_2 Z_4= Z_3 T & Z_3 Z_5=Z_4 (T-\alpha)\\
Z_2 Z_5 =T (T-\alpha) & Z_1 Z_4=T (T+\beta) & 
\end{array}
\end{equation}
so that the equations for the deformed geometry (\ref{def}) are
correctly reproduced also in field theory, using the identifications:
\begin{equation}
\alpha=-S_1 \qquad \beta=S_2
\end{equation} 

The same geometry has to be found using equivalent rank distributions,
for instance that reported in Figure \ref{distribution1}. We have
studied the corresponding field theory in the simpler case $P=0$, that
corresponds to a deformation with a single parameter $S_1=S_2$ (the
toric diagram is split into the sum of a triangle and a square). With
$N=1$, we have four non abelian $SU(M+1)$ gauge groups (faces
$3,4,6,7$). By performing a Seiberg duality with respect to groups $7$
and $4$ we
are left with only two non-abelian gauge groups and one can repeat easily 
the computation of the quantum modified moduli space.
We have checked again that the deformed geometry is correctly reproduced.

\section{Conclusions}

In this paper we have provided a simple method to compute anomaly free rank
distributions in quiver gauge theories corresponding to fractional
deformation branes or to $\mathcal N=2$ fractional branes. More
generally we have suggested that an efficient qualitative 
understanding of the IR
behavior of the gauge theory with fractional branes can be obtained by
looking at the weights $b_i$ associated with external legs in the
(p,q) web.

Note however that according to our proposal, and as already noted in
\cite{susyb1}, deformation branes and $\mathcal N=2$
fractional branes correspond to very special weights distributions
for the legs of the (p,q) web and moreover they can appear only when   
the toric diagram satisfies certain conditions. More general
distributions should lead to what we have called supersymmetry
breaking behavior: 
for toric quiver gauge theories there have been found only examples of 
runaway behavior \cite{susyb1,susyb2,susyb3,runaway,force}, but it would be
interesting to know whether this is a general feature of this class
of fractional branes or whether one can find cases with a meta-stable
vacuum.
 
When the toric singularity can be smoothed by a complex deformation,
we have seen that the gauge theory has a supersymmetric confining
vacuum when no regular branes remain in the IR.
We verified in a concrete example that the moduli space probed by a
regular brane, when we use our rule for finding rank distributions of
deformation branes, reproduces the deformed geometry: deformation
parameters correspond to gaugino condensates.
 We also pointed
out that the $\Psi$-map is a useful tool in performing these
computations and can explain in the general superconformal case why
the moduli space of the gauge theory built with the Fast Inverse
Algorithm matches with the toric geometry description. But a more general
understanding of the deformed case is required.

Another important problem to be further investigated is the existence
and the behavior of cascades for the various classes of fractional
branes. 
Many examples are known in the literature
\cite{multiflux,cascateypq,force}; at least for fractional deformation
branes and SB branes there seems to exist a cascade of Seiberg
dualities that, after passing through a certain number of possibly 
different phases of the gauge theory, sends the dimer back to
itself but with a decreased number of regular branes. However a
general study of cascades
requires a better understanding of the (toric) phases of the quiver
gauge theory.

\vspace{2em} 
\noindent {\Large{\bf Acknowledgments}}

\vspace{0.5em}

I would like to thank in primis Alberto Zaffaroni, and Andrea Brini,
Davide Forcella, Amihay Hanany for useful discussions and kind
encouragement.  
This work is supported in part by INFN and MURST under 
contract 2005-024045-004 and by 
the European Community's Human Potential Programme
MRTN-CT-2004-005104.

\end{document}